\documentclass[reqno,10pt,a4paper,table]{amsart}

\numberwithin{equation}{section}
\allowdisplaybreaks 

\usepackage{mathtools} 
\usepackage{amssymb,eucal,mathrsfs,setspace}
\usepackage[noadjust]{cite}
\usepackage{graphicx}
\usepackage{caption} 
\usepackage{collectbox}
\usepackage{array}
\newcolumntype{C}{>{$}c<{$}} 

\usepackage[largesc,theoremfont]{newtxtext}       
\usepackage[libertine,cmbraces,varbb]{newtxmath}  
\begingroup
  \makeatletter
  \@for\theoremstyle:=definition,remark,plain\do{%
    \expandafter\g@addto@macro\csname th@\theoremstyle\endcsname{%
      \addtolength\thm@preskip{.5\baselineskip plus .2\baselineskip minus .2\baselineskip}
      \addtolength\thm@postskip{.5\baselineskip plus .2\baselineskip minus .2\baselineskip}
    }%
  }
\endgroup

\usepackage[inner=24mm, outer=24mm, top=25mm, bottom=25mm, head=10mm, foot=10mm]{geometry}

\usepackage{enumitem}
\setlist[itemize]{leftmargin=*}               
\setlist[enumerate]{leftmargin=*,             
                    label=\textup{(\roman*)},
                    ref=\textup{\roman*}}     

\usepackage[colorlinks=true,citecolor=red,linkcolor=blue]{hyperref} 

\usepackage[capitalise,noabbrev]{cleveref} 

\usepackage{tikz}
\tikzset{nom/.style={circle,inner sep=2pt,fill=black!20!,minimum size=25pt}}

\newcommand{\commutant}[2]{\operatorname{Comm}\left(#1,#2\right)}

\renewcommand{\cong}{\simeq}
\renewcommand{\ncong}{\not\simeq}

\DeclareMathOperator{\tdeg}{t-deg} 

\renewcommand{\ge}{\geqslant} 
\renewcommand{\le}{\leqslant} 

\newcommand{\cc}{\mathsf{c}}   
\newcommand{\dd}{\mathrm{d}}   
\newcommand{\ii}{\mathfrak{i}}   
\newcommand{\kk}{\mathsf{k}}   
\newcommand{\qq}{\mathsf{q}}   
\newcommand{\zz}{\mathsf{z}}    

\DeclarePairedDelimiter{\brac}{\lparen}{\rparen}   
\DeclarePairedDelimiter{\sqbrac}{\lbrack}{\rbrack} 
\DeclarePairedDelimiter{\set}{\lbrace}{\rbrace}
\DeclarePairedDelimiter{\abs}{\lvert}{\rvert}

\DeclarePairedDelimiter{\ang}{\langle}{\rangle}    

\DeclarePairedDelimiter{\normord}{{:}}{{:}}        				

\DeclarePairedDelimiterX{\comm}[2]{\lbrack}{\rbrack}{#1 , #2}  
\DeclarePairedDelimiterX{\acomm}[2]{\lbrace}{\rbrace}{#1 , #2} 
\DeclarePairedDelimiterX{\inner}[2]{\langle}{\rangle}{#1 , #2} 
\DeclarePairedDelimiterX{\super}[2]{\lparen}{\rparen}{#1 \delimsize\vert \mathopen{} #2} 

\newcommand{\Ra}{\Rightarrow}
\newcommand{\lra}{\longrightarrow}
\newcommand{\ira}{\hookrightarrow}    
\newcommand{\lira}{\ensuremath{\lhook\joinrel\relbar\joinrel\rightarrow}} 

\newcommand{\dses}[3]{0 \lra #1 \lra #2 \lra #3 \lra 0} 

\newcommand{\ressymb}{\big\downarrow}
\newcommand{\indsymb}{\big\uparrow}
\newcommand{\res}[1]{#1 \mbox{$\ressymb$} {}}  
\newcommand{\ind}[1]{#1 \mbox{$\indsymb$} {}}  
\newcommand{\resind}[1]{\res{\ind{#1}}}

\let\hom\relax
\DeclareMathOperator{\hom}{Hom}

\newcommand{\VOA}[1]{\mathsf{#1}}    
\newcommand{\categ}[1]{\mathscr{#1}} 

\newcommand{\fld}[1]{\mathbb{#1}}    
\newcommand{\ZZ}{\fld{Z}}

\newcommand{\CC}{\fld{C}}

\newcommand{\wun}{\vvmathbb{1}}  

\newcommand{\klcat}{\categ{K\!\!L}}              

\newcommand{\vmodel}[1]{\VOA{M}(#1)}             
\newcommand{\slmodel}[1]{\VOA{A}_1(#1)}			     
\newcommand{\ospmodel}[1]{\VOA{B}_{0\vert1}(#1)} 
\newcommand{\svmodel}[1]{\VOA{SM}(#1)}           

\newcommand{\grp}[1]{\mathsf{#1}}              
\newcommand{\SLG}[2]{\grp{#1}_{#2}}            

\newcommand{\alg}[1]{\mathfrak{#1}}                      
\newcommand{\finite}[1]{#1}                              
\newcommand{\affine}[1]{\widehat{#1}}

\newcommand{\SLA}[2]{\finite{\alg{#1}}_{#2}}             
\newcommand{\SLSA}[3]{\finite{\alg{#1}} \super{#2}{#3}}  
\newcommand{\AKMA}[2]{\affine{\alg{#1}}_{#2}}            
\newcommand{\AKMSA}[3]{\affine{\alg{#1}} \super{#2}{#3}} 

\newcommand{\slenvalg}{\mathsf{U}(\SLA{sl}{2})}
\newcommand{\oenvalg}{\mathsf{U}(\SLSA{osp}{1}{2})} 
\newcommand{\aoenvalgk}{\affine{\mathsf{U}}_{\kk}}  
\newcommand{\ideal}[1]{\ang[\big]{#1}}

\newcommand{\sltwo}{\SLA{sl}{2}}
\newcommand{\asltwo}{\AKMA{sl}{2}}
\newcommand{\osp}{\SLSA{osp}{1}{2}}
\newcommand{\aosp}{\AKMSA{osp}{1}{2}}

\newcommand{\cas}{\Omega}                       
\newcommand{\scas}{\varsigma}                   

\newcommand{\supv}{\textup{Vir}}
\newcommand{\supsl}{\textup{sl}}
\newcommand{\suposp}{\textup{osp}}

\newcommand{\Mod}[1]{\mathcal{#1}}           

\newcommand{\vSymb}{\Mod{V}}       
\newcommand{\virr}[1]{\vSymb_{#1}} 

\newcommand{\slFinSymb}{\Mod{L}}    
\newcommand{\slDisSymb}{\Mod{D}}    
\newcommand{\slRelSymb}{\Mod{E}}    
\newcommand{\slProjSymb}{\Mod{S}}   

\newcommand{\aslfin}[1]{\slFinSymb_{#1}}         
\newcommand{\asldisp}[1]{\slDisSymb^+_{#1}}         
\newcommand{\asldism}[1]{\slDisSymb^-_{#1}}         
\newcommand{\asldispm}[1]{\slDisSymb^{\pm}_{#1}}         
\newcommand{\asldismp}[1]{\slDisSymb^{\mp}_{#1}}         

\newcommand{\aslrel}[2]{\slRelSymb_{#1; #2}}     
\newcommand{\aslrelredpm}[1]{\slRelSymb^{\pm}_{#1}}     

\newcommand{\aslproj}[1]{\slProjSymb_{#1}} 

\newcommand{\NS}{\textup{\textsc{NS}}}
\newcommand{\R}{\textup{\textsc{R}}}

\newcommand{\ospFinSymb}{\Mod{A}}    
\newcommand{\ospDisSymb}{\Mod{B}}    
\newcommand{\ospRelSymb}{\Mod{C}}    
\newcommand{\ospProjSymb}{\Mod{P}}   

\newcommand{\aospfin}[1]{\ospFinSymb_{#1}}                 
\newcommand{\aospdisp}[1]{\ospDisSymb^+_{#1}}              
\newcommand{\aospdism}[1]{\ospDisSymb^-_{#1}}              
\newcommand{\aospdispm}[1]{\ospDisSymb^{\pm}_{#1}}         
\newcommand{\aospdismp}[1]{\ospDisSymb^{\mp}_{#1}}         

\newcommand{\aosprel}[2]{\ospRelSymb_{#1; #2}}             
\newcommand{\aosprelredpm}[1]{\ospRelSymb^{\pm}_{#1}}      

\newcommand{\aospproj}[1]{\ospProjSymb_{#1}} 

\newcommand{\aospfintemp}[1]{\widetilde{\ospFinSymb}_{#1}} 

\newcommand{\parrev}{\Pi}  			
\newcommand{\sfsymb}{\sigma} 	 	
\newcommand{\sfsymbsl}{\sfsymb_{\supsl}} 	  	 
\newcommand{\sfsymbosp}{\sfsymb_{\suposp}} 	 	 
\newcommand{\sfmod}[2]{\sfsymb^{#2}(#1)}       
\newcommand{\sfmodsl}[2]{\sfsymbsl^{#2}(#1)}	 
\newcommand{\sfmodosp}[2]{\sfsymbosp^{#2}(#1)} 

\DeclareMathOperator{\tr}{tr}

\newcommand{\jth}[1]{\vartheta_{#1}}
\newcommand{\fjth}[2]{\jth{#1} \brac*{#2}}

\newcommand{\Gr}[1]{\sqbrac[\big]{#1}}                 
\newcommand{\tGr}[1]{\sqbrac{#1}}                      

\newcommand{\traceover}[1]{\tr_{\raisebox{-2pt}{$\scriptstyle #1$}}} 

\DeclareMathOperator{\chmap}{ch}
\DeclareMathOperator{\schmap}{sch}

\newcommand{\ch}[1]{\chmap \Gr{#1}}                    
\newcommand{\fch}[1]{\ch{#1} \brac[\big]{\zz;\qq}}          
\newcommand{\fchsf}[1]{\ch{#1} \brac[\big]{\zz\qq^{\ell/2};\qq}} 

\newcommand{\chvir}[2]{{\chi^{#2}_{#1}}}	         
\newcommand{\fchvir}[2]{{\chvir{#1}{#2}(\qq)}}	   

\newcommand{\chsvir}[2]{{\psi^{#2}_{#1}}}	             
\newcommand{\fchsvir}[2]{{\chsvir{#1}{#2}(\qq)}}	     
\newcommand{\schsvir}[2]{{\widetilde{\psi}^{#2}_{#1}}} 
\newcommand{\fschsvir}[2]{{\schsvir{#1}{#2}(\qq)}}     

\newcommand{\sch}[1]{\schmap \Gr{#1}}                     
\newcommand{\fsch}[1]{\sch{#1}(\zz;\qq)}          

\newcommand{\fuse}{\mathbin{\times}}                                      
\newcommand{\Grfuse}{\mathbin{\boxtimes}}                            
\newcommand{\vcoe}[3]{\mathsf{N}_{#2}^{(#1) \: #3}}               

\newcommand{\zhu}[2]{\mathrm{Zhu}^{#1} \sqbrac*{#2}} 
\newcommand{\nszhu}[1]{\zhu{\NS}{#1}}
\newcommand{\rzhu}[1]{\zhu{\R}{#1}}

\theoremstyle{plain}
\newtheorem{theorem}{Theorem}

\newtheorem{proposition}[theorem]{Proposition}
\newtheorem{conjecture}{Conjecture}

\Crefname{conjecture}{Conjecture}{Conjectures} 

\newcommand{\cft}{conformal field theory}
\newcommand{\cfts}{conformal field theories}

\newcommand{\voa}{vertex operator algebra}
\newcommand{\voas}{vertex operator algebras}
\newcommand{\vosa}{vertex operator subalgebra}

\newcommand{\svoa}{vertex operator superalgebra}
\newcommand{\svoas}{vertex operator superalgebras}

\newcommand{\opes}{operator product expansions}
\newcommand{\uea}{universal enveloping algebra}

\newcommand{\hw}{highest-weight}
\newcommand{\hwv}{\hw{} state} 
\newcommand{\hwvs}{\hw{} states} 
\newcommand{\hwm}{\hw{} module}
\newcommand{\hwms}{\hw{} modules}
\newcommand{\sv}{singular vector}

\newcommand{\rhs}{right-hand side}

\newcommand{\ns}{Neveu-Schwarz}

\makeatletter
\renewcommand\author@andify{%
  \nxandlist {\unskip ,\penalty-1 \space\ignorespaces}%
    {\unskip {} \@@and~}%
    {\unskip \penalty-2 \space \@@and~}%
}
\makeatother

\begin{document}

\title[]{Cosets, characters and fusion for \\ admissible-level $\mathfrak{osp}(1 \vert 2)$ minimal models.}

\author{Thomas Creutzig}
\address{
Department of Mathematical and Statistical Sciences \\
University of Alberta\\
Edmonton, Canada T6G~2G1 and  \\
Research Institute for Mathematical Sciences \\ Kyoto University\\ Kyoto Japan 606-8502.
}
\email{creutzig@ualberta.ca}

\author[S~Kanade]{Shashank Kanade}
\address[Shashank Kanade]{
Department of Mathematics \\
University of Denver \\
Denver, USA, 80208.
}
\email{shashank.kanade@du.edu}

\author[T~Liu]{Tianshu Liu}
\address[Tianshu Liu]{
School of Mathematics and Statistics \\
University of Melbourne \\
Parkville, Australia, 3010.
}
\email{tianshul@student.unimelb.edu.au}

\author[D~Ridout]{David Ridout}
\address[David Ridout]{
School of Mathematics and Statistics \\
University of Melbourne \\
Parkville, Australia, 3010.
}
\email{david.ridout@unimelb.edu.au}


\begin{abstract}
	We study the minimal models associated to $\mathfrak{osp}(1 \vert 2)$, otherwise known as the fractional-level Wess-Zumino-Witten models of $\mathfrak{osp}(1 \vert 2)$. Since these minimal models are extensions of the tensor product of certain Virasoro and $\mathfrak{sl}_2$ minimal models, we can induce the known structures of the representations of the latter models to get a rather complete understanding of the minimal models of $\mathfrak{osp}(1 \vert 2)$. In particular, we classify the irreducible relaxed highest-weight modules, determine their characters and compute their Grothendieck fusion rules. We also discuss conjectures for their (genuine) fusion products and the projective covers of the irreducibles.
\end{abstract}

\maketitle

\onehalfspacing

\section{Introduction} \label{sec:Intro}

This project is part of a programme to understand the admissible-level Wess-Zumino-Witten (WZW) models for a Lie algebra or superalgebra $\alg{g}$. While the theories with non-negative integer levels and simple Lie algebras lead to rational conformal field theories and, as such, are very well understood, the situation is much more complicated and rich for other levels or when superalgebras are involved. Indeed, the non-rational admissible-level WZW models are expected to be prime examples of logarithmic conformal field theories, these being models that admit representations on which the hamiltonian acts non-diagonalisably, leading to correlation functions with logarithmic singularities. Another interesting feature of these models is that they have a continuous spectrum of modules.

We view our programme as complementary to older approaches.  In particular, Quella, Saleur, Schomerus \emph{et al.} \cite{RozQua92,GotRep07,SalGL106,GotWZN07,SalSU207,QueFre07,CreGL07,CreGL08,CreRo08} approached supergroup WZW theories via free field realisations and semiclassical limits (the minisuperspace analysis), the interest being rather in features of the WZW theory of the supergroup at integer levels.  Another approach employed was to learn more about the conformal field theory using the mock modular behaviour of certain irreducible characters \cite{SemHig05,JohMod00,AlfMoc14}.  The relatively accessible case of $\alg{g} = \SLSA{gl}{1}{1}$ has also been studied from a more algebraic perspective by two of us \cite{CreRel11,CreGL11}.

Presently, we have a very good picture in the case of $\alg{g} = \SLA{sl}{2}$ \cite{AdaVer95,DonVer97,GabFus01, RidSL208,RidSL210,RidFus10,CreMod12,CreCos13,CreMod13}.  In order to extend our understanding to more sophisticated theories, one has to develop some basic strategies. First, one has to study the general theory of relaxed \hwms.  These natural generalisations of the usual \hwms{} were introduced in the \cft{} literature in \cite{FeiEqu98} for $\alg{g} = \SLA{sl}{2}$, though they had already appeared in mathematics classifications such as \cite{AdaVer95}, but have only recently been formalised in a general setting \cite{RidRel15}.  Since then, the role played by irreducible relaxed \hwms{} in facilitating the study of general admissible-level WZW models has been widely appreciated and the field has been rapidly developing, see \cite{AraWei16,AdaRea17,KawRel18,KawRel18b} for example.

Second, one should develop techniques to reconstruct, at least in favourable circumstances, the representation theory of the algebra of interest in terms of those of subalgebras. We call this technique the (inverse) coset construction. This formalism has recently been developed in detail and rigour in \cite{HuaBra15,CreSim15,CreSch16,CreTen17} and, as a preparatory example, we have studied the logarithmic parafermion algebras of $\SLA{sl}{2}$ at (negative) admissible levels \cite{AugMod17}. The present paper is concerned with the minimal models for $\alg{g} = \osp$, these being the admissible-level WZW models, building on the insights obtained for a particular level in \cite{RidAdm17}.  In a sequel, the results of this paper will be combined with those of \cite{AugMod17} in order to understand the minimal models of $\SLSA{sl}{2}{1}$ at admissible levels.

Recently, vertex superalgebras and their modules have appeared as invariants of four-dimensional superconformal theories. For example, those associated with $\SLA{sl}{2}$ and certain subregular W-algebras at various admissible levels arise in the study of Argyres-Douglas theories \cite{BuiAD15i,CorSchur15,CreAD17}. Their vacuum characters coincide with the Schur indices of these four-dimensional theories, while the indices of line defects are identified with characters of \hwms{} and those of surface indices seem to correspond to relaxed \hw{} characters \cite{CorSur17}.

Further examples include topological twisted four-dimensional supersymmetric gauge theories.  There, \svoas{} appear at the junction of three-dimensional topological boundary conditions. Categories of line defects ending on boundaries at which these conditions are imposed correspond to subcategories of modules of the junction \svoa. The best understood example is the level-$1$ affine \svoa{} of the exceptional simple Lie superalgebra $\SLSA{d}{2}{1;\alpha}$ which appears as a certain junction subalgebra in $SU(2)$ gauge theory \cite{CreS17}. But, an $\osp$ \svoa{} also appears, specifically as a junction of the so-called $D_{0,1}$ and $N_{2,1}$ boundary conditions \cite{CrS18}. In fact, the coset studied here (see \eqref{coset} below) has an interpretation in gauge theory as the junction for $\osp$ being obtained by concatenating the junctions corresponding to $\sltwo$ and Virasoro. As above, the categories of line defects correspond to ordinary modules and spectrally flown images of the vacuum, while one expects that relaxed \hwms{} correspond to categories of surface defects.

\subsection{The inverse coset construction}

Our strategy in this article is to invert the coset construction. This is a rather subtle story and needs a little bit of vertex algebra tensor category theory.  We aim to understand the conformal field theory of $\osp$ at admissible level, whose symmetry algebra we denote by $\ospmodel{p,v}$. The coset of this theory corresponding to the $\SLA{sl}{2}$ subtheory $\slmodel{u,v}$ is a rational Virasoro minimal model $\vmodel{p,u}$:
\begin{equation}
	\vmodel{p,u} \cong \commutant{\slmodel{u,v}}{\ospmodel{p,v}} \equiv \frac{\ospmodel{p,v}}{\slmodel{u,v}} \qquad \text{($2u=p+v$).}\label{coset}
\end{equation}
This means that every module of $\ospmodel{p,v}$ is also a module of the tensor product of the two subalgebras $\slmodel{u,v}$ and $\vmodel{p,u}$. We thus want to construct the representations of $\ospmodel{p,v}$ from the known ones of these subalgebras. The mathematical tool that accomplishes this is \emph{induction}.

In vertex algebra language, the bigger algebra $\ospmodel{p,v}$ is a commutative superalgebra object in the category of modules for the small algebra $\slmodel{u,v}\otimes \vmodel{p,u}$ \cite{CreSim15}. Moreover, there is a notion of local (and Ramond-twisted) superalgebra modules and these are exactly the Neveu-Schwarz (and Ramond) modules of $\ospmodel{p,v}$ \cite{HuaBra15, CreTen17}. Locality here means that the operator product algebra with the currents of $\osp$ is monodromy-free. Our task is thus to find all these local (and Ramond-twisted) modules. Another result of \cite{CreTen17} is that induction is a vertex tensor functor from a subcategory of modules for the smaller algebra to this category of local modules. The objects of this subcategory are exactly those that satisfy a certain locality condition that can be rephrased in terms of conformal dimensions, giving us a clear procedure to search for, and identify, these modules. Even better, the induction functor is monoidal \cite{CreTen17} and hence it preserves the fusion rules, so we can easily compute the $\ospmodel{p,v}$ fusion rules from those known for $\slmodel{u,v}$ \cite{GabFus01,RidFus10,CreMod12,CreMod13,CreBra17} and $\vmodel{p,u}$ \cite{BelInf84,FreVer92}.

On physical grounds, conformal field theory is always expected to require a vertex tensor category in the sense of Huang-Lepowsky-Zhang \cite{HuaLog10} and so one expects that an appropriate version of Verlinde's formula holds. Verifying the existence of a vertex tensor category structure and proving a Verlinde formula for non-rational \voas{} are two of the deepest problems in vertex algebra theory. In our case, both have recently been proven for the subcategory of ordinary modules of $\slmodel{u,v}$ \cite{CreBra17} (and all other simply-laced Lie algebras \cite{CreBra17,CreTen18}) so that the results reported here are completely rigorous within this subcategory.  In general, our results depend on the conjectural Verlinde formula for $\slmodel{u,v}$ of \cite{CreMod12,CreMod13}, developed in \cite{CreRel11}, and the conjectural existence of a vertex tensor category structure on the $\slmodel{u,v}$-modules.

\subsection{Outline and Results}

We start in \cref{sec:Back} with the necessary background, meaning that we introduce the Virasoro and $\SLA{sl}{2}$ minimal models and fix their notation.  These are the building blocks of the $\osp$ minimal models which we set up in \cref{sec:osp}. Next, in \cref{sec:coset}, we explain the realisation of each minimal model of $\osp$ in terms of a \svoa{} extension of the tensor product of certain Virasoro minimal model with an $\SLA{sl}{2}$ one.  In particular, we review and explain the ``character'' proof of the coset \eqref{coset} presented in \cite{CreRep17}.  In order to deduce various facets of the representation theory of the $\osp$ minimal models, we also have to explain some basic properties of the theory of vertex algebra extensions using the language of induction and restriction. This is done in \cref{sec:vtcs}.

With this setup, we are now able to construct modules of the $\osp$ minimal models via induction and so we start \cref{sec:construct} by finding all modules of the tensor product \vosa{} that induce to irreducible \ns{} and Ramond modules over $\aosp$. We then identify these modules by determining their global parities and other characterising data for $\osp$ (highest weights, conformal dimensions and super-Casimir eigenvalues). Moreover, the construction makes it easy to explicitly state the characters and supercharacters of the induced representations.

It is expected, but is \emph{a priori} not clear, that one gets all irreducible modules of the $\osp$ minimal models via induction. In \cref{sec:complete}, we prove that this is so, for relaxed \hwms{}, by combining the information we get from the explicit constructions with some simple observations concerning Zhu's algebra.  This provides a new, and relatively straightforward, proof of the recent classification \cite{WooAdm18} of Wood.

Finally, we use the fact that fusion respects induction to immediately deduce (conjectural) fusion rules for the irreducible modules of the $\osp$ minimal models, see \cref{sec:Fusion}.  Presently, even in the case of the $\sltwo$ minimal models, the projective covers of the irreducible modules are not known. We use this opportunity to list our conjectures for their structures in \cref{app:frsl2}, explaining that they are consistent with the general expectations for fusion rules in rigid tensor categories. This allows us to construct (conjectured) projectives for the $\osp$ minimal models and state some of their fusion rules (\cref{sec:projective}).

\section*{Acknowledgements}

TC is supported by the Natural Sciences and Engineering Research Council of Canada (RES0020460).
SK acknowledges the support of an Endeavour Research Fellowship (2017), offered by the Australian Government's Department of Education and Training, and a PIMS post-doctoral fellowship. SK is presently supported by a start-up grant provided by University of Denver.
TL's research is supported by a University Research Scholarship from the University of Melbourne.
DR's research is supported by the Australian Research Council Discovery Project DP160101520 and the Australian Research Council Centre of Excellence for Mathematical and Statistical Frontiers CE140100049.

\section{Background} \label{sec:Back}

\subsection{Conventions} \label{sec:voas}

\subsubsection{Virasoro minimal models} \label{sec:vir}

A minimal model \cite{BelInf84} of the Virasoro algebra $\alg{Vir}$
\begin{equation}
\comm{L^{\supv}_m}{L^{\supv}_n}=(m-n)L^{\supv}_{m+n}+\frac{m^3-m}{12}\delta_{m+n,0}\cc^{\supv}
\end{equation}
is denoted by $\vmodel{p,u}$, where $p,u \ge 2$ are coprime integers parametrising the central charge
\begin{equation}
\cc^{\supv}=1-6\frac{(u-p)^2}{pu}.
\end{equation}
The irreducible $\vmodel{p,u}$-modules are the \hw{} $\alg{Vir}$-modules $\virr{r,s}$, where $1\le r \le p-1$ and $1\le s \le u-1$, whose \hwvs{} have conformal dimension
\begin{equation} \label{eq:vconfdim}
\Delta^{\supv}_{r,s}=\frac{(ur-ps)^2-(u-p)^2}{4pu}.
\end{equation}
Note that $\virr{r,s}=\virr{p-r,u-s}$.

The character of the irreducible $\vmodel{p,u}$-module $\virr{r,s}$ is given by
\begin{equation} \label{eq:chvir}
\fchvir{r,s}{p,u}=\tr_{\virr{r,s}}\qq^{L^{\supv}_0-\cc^{\supv}/24}=\frac{1}{\eta(\qq)} \sum_{n\in\ZZ} \sqbrac[\big]{\qq^{(2pun+ur-ps)^2/4pu} - \qq^{(2pun+ur+ps)^2/4pu}},
\end{equation}
where $\eta(\qq)$ is the Dedekind eta function. The minimal model $\vmodel{p,u}$ is rational \cite{WanRat93,RidJac14} and the fusion rules are
\begin{equation} \label{eq:frvir}
\virr{r,s}\fuse\virr{r',s'}\cong\bigoplus_{r''=1}^{p-1} \bigoplus_{s''=1}^{u-1} \vcoe{p,u}{(r,s), (r',s')}{(r'',s'')} \, \virr{r'',s''},
\end{equation}
where $\vcoe{p,u}{(r,s), (r',s')}{(r'',s'')} = \vcoe{p}{r, r'}{r''} \vcoe{u}{s, s'}{s''} $ and
\begin{equation} \label{vircoe}
\vcoe{t}{i,j}{k}=
\begin{cases*}
1, & if $\abs{i-j}+1\le k\le \min \set{i+j-1,2t-i-j-1}$ and $i+j+k$ is odd,\\
0, & otherwise.
\end{cases*}
\end{equation}
We note that $\virr{1,1} = \virr{p-1,u-1}$ is the vacuum module and that when $p$ and $u$ are both greater than $2$, $\virr{p-1,1} = \virr{1,u-1}$ is a (distinct) simple current of order $2$:  $\virr{p-1,1} \fuse \virr{p-1,1} \cong \virr{1,1}$.

\subsubsection{$\sltwo$ minimal models} \label{sec:sl2}

The affine Kac-Moody algebra $\asltwo$ has a standard basis in which the non-zero commutation relations are
\begin{equation} \label{eq:commsl2}
	\comm{h_m}{e_n} = +2 e_{m+n}, \quad \comm{h_m}{h_n} = 2m \delta_{m+n,0} \kk, \quad
	\comm{e_m}{f_n} = h_{m+n} + m \delta_{m+n,0} \kk, \quad \comm{h_m}{f_n} = -2 f_{m+n}.
\end{equation}
The universal affine vertex algebra of level $\kk$ associated to $\asltwo$ is not simple when \cite{GorSim07}
\begin{equation} \label{eq:admksl2}
	\kk+2=\frac{u}{v}, \quad u\in\ZZ_{\ge2},\ v\in\ZZ_{\ge1},\ \gcd \set{u,v}=1.
\end{equation}
Its simple quotient is referred to as the level-$\kk$ $\sltwo$ minimal model and will be denoted by $\slmodel{u,v}$. The energy-momentum tensor of this minimal model is given by the Sugawara construction \cite{SugFie68,SegUni81} as
\begin{equation}
	T^{\supsl}(z)=\frac{1}{2(\kk+2)}\sqbrac*{\frac 1 2 \normord{hh}(z)+\normord{ef}(z)+\normord{fe}(z)}
\end{equation}
and the central charge of $\slmodel{u,v}$ is
\begin{equation}
\cc^{\supsl}=3-\frac{6v}{u}.
\end{equation}

The generators of $\asltwo$ admit a number of automorphisms including \cite{SchCom87} spectral flow $\sfsymbsl^{\ell}$, where $\ell \in \ZZ$, which preserves the level $\kk$ and acts on the other generators by
\begin{equation}
\sfmodsl{e_n}{\ell}=e_{n-\ell}, \quad \sfmodsl{h_n}{\ell}=h_n-\delta_{n,0}\ell\kk, \quad \sfmodsl{f_n}{\ell}=f_{n+\ell}.
\end{equation}
The zero mode $L^{\supsl}_0$ of the energy momentum tensor, whose eigenvalue is the conformal dimension, satisfies
\begin{equation}
\sfmodsl{L^{\supsl}_0}{\ell}=L^{\supsl}_0-\frac 1 2 \ell h_0+\frac 1 4 \ell^2\kk.
\end{equation}
Spectral flow also acts on $\slmodel{u,v}$-modules through composition with the corresponding representations.  We shall denote the spectral flows of such a module $\Mod{M}$ by $\sfmodsl{\Mod{M}}{\ell}$.

The minimal model $\slmodel{u,v}$ is unitary when $v=1$, in which case the level $\kk$ is a non-negative integer. The minimal model $\slmodel{u,1}$ is rational \cite{WitNon84}, so has a finite number of irreducible modules $\aslfin{r,0}$, where $1\le r \le u-1$, which happen to be integrable and \hw{}. The $h_0$-charge and conformal dimension of the \hwv{} of $\aslfin{r,0}$ are given by
\begin{equation} \label{eq:usl2weights}
\lambda^{\supsl}_{r,0}=r-1 \quad \text{and} \quad \Delta^{\supsl}_{r,0}=\frac{r^2-1}{4u},
\end{equation}
respectively.  The spectral flows of these irreducibles satisfy
\begin{equation}
	\sfmodsl{\aslfin{r,0}}{} \cong \aslfin{u-r,0}
\end{equation}
and their characters are given by
\begin{equation}
	\fch{\aslfin{r,0}}=\traceover{\aslfin{r,0}} \zz^{h_0} \qq^{L^{\supsl}_0-\cc^{\supsl}/24}
	=\frac{\qq^{\Delta^{\supsl}_{r,0}-\cc^{\supsl}/24+1/8}}{\ii \fjth{1}{\zz^2;\qq}}\sum_{j\in\ZZ}\left(\zz^{2uj+r}-\zz^{-2uj-r}\right) \qq^{j(uj+r)}, \label{slcharuni}
\end{equation}
where $\jth{1}$ denotes a Jacobi theta function (see \cite{RidSL208} for our conventions).  Finally, the fusion rules are given by
\begin{equation} \label{eq:ufrsl2}
	\aslfin{r,0} \fuse \aslfin{r',0} \cong \bigoplus_{r''=1}^{u-1} \vcoe{u}{r,r'}{r''} \, \aslfin{r'',0}.
\end{equation}
The vacuum module is $\aslfin{1,0}$ and, for $u>2$, $\aslfin{u-1,0}$ is a simple current of order $2$.  The $\slmodel{u,1}$ \cfts{} are commonly known as the WZW models on the Lie group $\SLG{SU}{2}$.

When $v\neq1$, the minimal model $\slmodel{u,v}$ is non-unitary and logarithmic \cite{RidRel15} with fractional level $k=\frac{u}{v}-2\notin\ZZ$. In this case, we generalise the parametrisation of the $h_0$-charge and conformal dimensions from \eqref{eq:usl2weights} to
\begin{equation} \label{eq:gensl2wts}
	\lambda^{\supsl}_{r,s} = r-1-\frac{u}{v}s, \quad \Delta^{\supsl}_{r,s} = \frac{(vr-us)^2-v^2}{4uv}.
\end{equation}
With this, the irreducible $\slmodel{u,v}$-modules come in several different classes, including those in the following list:
\begin{itemize}
	\item The $\aslfin{r,0}$, where $1\le r\le u-1$. Each is an irreducible \hwm{} whose space of ground states is finite-dimensional. The \hwv{} of each module has $h_0$-charge $\lambda^{\supsl}_{r,0}$ and conformal dimension $\Delta^{\supsl}_{r,0}$.
	\item The $\asldisp{r,s}$, where $1\le r\le u-1$ and $1\le s\le v-1$. Each is an irreducible \hwm{} whose \hwv{} has charge $\lambda^{\supsl}_{r,s}$ and conformal dimension $\Delta^{\supsl}_{r,s}$. The space of ground states forms an irreducible infinite-dimensional Verma module for the horizontal subalgebra $\sltwo$.
	\item The $\asldism{r,s}$, where $1\le r\le u-1$ and $1\le s\le v-1$. These are defined to be the conjugates of the $\asldisp{r,s}$, meaning that $\asldism{r,s}$ is obtained from $\asldisp{r,s}$ by twisting the $\slmodel{u,v}$-action by the Weyl reflection of $\sltwo$.  It follows that the ground states of the $\asldism{r,s}$ also have conformal dimension $\Delta^{\supsl}_{r,s}$.
	\item The $\aslrel{\lambda}{r,s}$, where $1\le r\le u-1$, $1\le s\le v-1$ and $\lambda \in \CC$ satisfy $\lambda\neq\lambda^{\supsl}_{r,s}, \lambda^{\supsl}_{u-r,v-s} \pmod{2}$.  Each is an irreducible relaxed \hwm{} whose ground states have $h_0$-charges equal to $\lambda \pmod{2}$ and conformal dimension $\Delta^{\supsl}_{r,s}$.
	\item Spectral flows of all of the irreducible modules above.  This generally gives new irreducibles, though there are some isomorphisms to note, in particular
	\begin{equation} \label{eq:sfsl2}
		\sfmodsl{\aslfin{r,0}}{\pm1} \cong \asldispm{u-r,v-1}, \qquad
		\sfmodsl{\asldispm{r,s}}{\mp1} \cong \asldismp{u-r,v-1-s} \quad \text{($s \neq v-1$).}
	\end{equation}
\end{itemize}
Apart from the spectral flows, this classification originally appeared in \cite{AdaVer95}.  More recent alternative proofs may be found in \cite{RidRel15,KawRel18b}.

There exist additional classes of irreducible $\slmodel{u,v}$-modules, for instance the Whittaker modules of \cite{AdaRea17}.  However, these are not expected to be needed for the construction of the corresponding (logarithmic) \cfts{}.  One does, however, need certain reducible but indecomposable $\slmodel{u,v}$-modules, in particular the relaxed \hwms{} $\aslrelredpm{r,s}$, where $1\le r\le u-1$ and $1\le s\le v-1$.  These have ground states whose $h_0$-charges are equal to $\lambda^{\supsl}_{r,s} \pmod{2}$ and whose conformal dimension is $\Delta^{\supsl}_{r,s}$.  Moreover, $\aslrelredpm{r,s}$ has a submodule isomorphic to $\asldispm{r,s}$ and its quotient by this submodule is isomorphic to $\asldismp{u-r,v-s}$.  In other words, the following sequence is exact and non-split:
\begin{equation} \label{eq:essl2}
	\dses{\asldispm{r,s}}{\aslrelredpm{r,s}}{\asldismp{u-r,v-s}}.
\end{equation}

The characters of the $\slmodel{u,v}$-modules introduced above are given by
\begin{subequations} \label{eq:chsl2}
\begin{align}
\fch{\aslfin{r,0}}&=\frac{\qq^{\Delta^{\supsl}_{r,0}-\cc^{\supsl}/24+1/8}}{\ii\vartheta_1(\zz^2;\qq)}\sum_{j\in\ZZ} \brac*{\zz^{2uj+r}-\zz^{-2uj-r}} \qq^{vj(uj+r)},\label{charL}\\
\fch{\asldispm{r,s}}&=\frac{\zz^{\pm(\lambda_{r,s}+1)} \qq^{\Delta^{\supsl}_{r,s}-\cc^{\supsl}/24+1/8}}{\pm\ii\vartheta_1(\zz^2;\qq)} \sum_{j\in\ZZ}\sqbrac*{\zz^{\pm2uj}\qq^{j(uvj+vr-us)}-\zz^{\pm2(uj-r)}\qq^{(uj-r)(vj-s)}},\label{charD}\\
\fch{\aslrel{\lambda}{r,s}}&=
\frac{\zz^{\lambda} \fchvir{r,s}{u,v}}{{\eta(\qq)}^2} \sum_{n\in\ZZ} \zz^{2n}, \qquad
\fch{\aslrelredpm{r,s}}=\frac{\zz^{\lambda_{r,s}} \fchvir{r,s}{u,v}}{{\eta(\qq)}^2} \sum_{n\in\ZZ} \zz^{2n}\label{charE}
\end{align}
\end{subequations}
where we recall that $\chvir{r,s}{u,v}$ in \eqref{charE} denotes the character of the irreducible $\vmodel{u,v}$-module $\virr{r,s}$.  The formula for the $\aslrel{\lambda}{r,s}$ was originally conjectured in \cite{CreMod13} and was proven in \cite{AdaRea17}, for generic values of the parameters, and in full generality in \cite{KawRel18}.  The characters of the spectral flows of an $\slmodel{u,v}$-module $\Mod{M}$ are easily obtained from
\begin{equation} \label{eq:sfcharsl2}
	\fch{\sfmodsl{\Mod{M}}{\ell}} = \zz^{\ell \kk} \qq^{\ell^2 \kk/4} \fchsf{\Mod{M}},
\end{equation}
though one should be careful with convergence regions (see \cite{RidSL208,CreMod13}).

The fusion rules of the irreducible $\slmodel{u,v}$-modules with $v \neq 1$ are only known for $(u,v) = (2,3)$ \cite{GabFus01,CreMod12} and $(u,v) = (3,2)$ \cite{RidFus10}, where they were computed using the Nahm-Gaberdiel-Kausch algorithm \cite{NahQua94,GabInd96}.  On the other hand, the Grothendieck fusion rules are known \cite{CreMod12,CreMod13}, under the twin conjectures that the Grothendieck fusion coefficients are well defined and that the standard Verlinde formula of \cite{CreLog13,RidVer14} computes them.  We list these rules in \cref{app:frsl2} for convenience.  Note that these conjectures imply the following fusion rules for general $u$ and $v$:
\begin{equation} \label{eq:frsl2Lx}
	\aslfin{r,0} \fuse \aslfin{r',0} \cong \bigoplus_{r''=1}^{u-1} \vcoe{u}{r,r'}{r''} \aslfin{r'',0}, \quad
	\aslfin{r,0} \fuse \asldispm{r',s'} \cong \bigoplus_{r''=1}^{u-1} \vcoe{u}{r,r'}{r''} \asldispm{r'',s'}, \quad
	\aslfin{r,0} \fuse \aslrel{\lambda}{r',s'} \cong \bigoplus_{r''=1}^{u-1} \vcoe{u}{r,r'}{r''} \aslrel{\lambda+\lambda_{r,0}}{r'',s'}.
\end{equation}
We emphasise that the fusion rules that decompose $\aslfin{r,0} \fuse \aslfin{r',0}$ have recently been proven in \cite{CreBra17}.  It follows that $\aslfin{1,0}$ and $\aslfin{u-1,0}$ are again the vacuum module and a simple current of order $2$ (if $u>2$), respectively.

\subsubsection{$\osp$ minimal models} \label{sec:osp}

The affine Kac-Moody superalgebra $\aosp$ is generated by bosonic modes $e_n$, $h_n$ and $f_n$, as well as fermionic modes $x_n$ and $y_n$.  Their non-zero (anti)commutation relations are given by \eqref{eq:commsl2} (the bosonic subalgebra of $\aosp$ is isomorphic to $\asltwo$) along with
\begin{equation}
	\begin{gathered}
		\comm{e_m}{y_s} = -x_{m+s}, \quad \comm{h_m}{x_s} = x_{m+s}, \quad \comm{h_m}{y_s} = -y_{m+s}, \quad \comm{f_m}{x_s} = -y_{m+s}, \\
		\acomm{x_r}{x_s} = 2e_{r+s}, \quad \acomm{x_r}{y_s} = h_{r+s} + 2r \delta_{r+s,0} \kk, \quad \acomm{y_r}{y_s} = -2 f_{r+s}.
	\end{gathered}
\end{equation}
There are actually two different versions of $\aosp$, one with $r,s \in \ZZ$ and another with $r,s \in \ZZ + \frac{1}{2}$.  Modules of the first version belong to the \ns{} sector, while those of the second belong to the Ramond sector.

The level-$\kk$ $\osp$ minimal model $\ospmodel{p,v}$ is defined to be the simple quotient of the universal \svoa{} associated to $\aosp$ with \cite{GorSim07}
\begin{equation} \label{eq:admkosp}
\kk=-\frac 3 2 +\frac{p}{2v}, \qquad p\in\ZZ_{\ge2},\ v\in\ZZ_{\ge1},\ \frac{p+v}{2} \in \ZZ,\ \gcd \set*{p,\frac{p+v}{2}}=1.
\end{equation}
The energy-momentum tensor provided by the Sugawara construction is
\begin{equation} \label{eq:emtosp}
T^{\suposp}(z)=\frac{1}{2\kk+3}\sqbrac*{\frac 1 2 \normord{hh}(z)+\normord{ef}(z)+\normord{fe}(z)-\frac{1}{2}\normord{xy}(z)+\frac{1}{2}\normord{yx}(z)}
\end{equation}
and the central charge is
\begin{equation}
\cc^{\suposp}= \frac{\kk}{2\kk+3} = 1-\frac{3v}{p}.
\end{equation}

Spectral flow acts on the generators of $\aosp$ and the Virasoro zero mode $L^{\suposp}_0$ obtained from \eqref{eq:emtosp} as follows:
\begin{equation}
\begin{gathered}
\sfmodosp{e_n}{\ell}=e_{n-\ell}, \qquad \sfmodosp{h_n}{\ell}=h_n-\delta_{n,0}\ell\kk, \qquad \sfmodosp{f_n}{\ell}=f_{n+\ell}\\
\sfmodosp{x_n}{\ell}=x_{n-\ell/2}, \qquad \sfmodosp{y_n}{\ell}=y_{n+\ell/2}\\
\sfmodosp{L^{\suposp}_0}{\ell}=L^{\suposp}_0-\frac 1 2 \ell h_0+\frac 1 4 \ell^2\kk.
\end{gathered}
\end{equation}
Note that restricting $\sfsymbosp$ to the bosonic subalgebra $\asltwo$ recovers $\sfsymbsl$.  As with $\slmodel{u,v}$-modules, the spectral flow $\sfmodosp{\Mod{M}}{\ell}$ of a $\ospmodel{p,v}$-module $\Mod{M}$ is another $\ospmodel{p,v}$-module.  If $\ell \in 2\ZZ$, then spectral flow preserves the sector (\ns{} or Ramond) of the module while these sectors are exchanged if $\ell \in 2\ZZ+1$.

The classification of irreducible relaxed \hw{} $\ospmodel{p,v}$-modules has only recently been completed in \cite{WooAdm18}, see also \cite{KawRel18b}.  Our aim here is to provide an alternative classification that relies on a coset construction.  This has the advantage that it will also allow us to easily deduce the characters, which were also only recently calculated \cite{KawRel18}, as well as the Grothendieck fusion rules, which were previously unknown.  To prepare for this classification and to fix notation, we introduce the irreducible relaxed \hw{} $\aosp$-modules following \cite{RidAdm17}:
\begin{itemize}
	\item The ${}^{\NS} \aospfin{\lambda}$ (${}^{\R} \aospfin{\lambda}$), where $\lambda \in \ZZ_{\ge 0}$.  Each is an irreducible \hwm{} in the \ns{} (Ramond) sector whose space of ground states forms an irreducible finite-dimensional module for $\osp$ ($\sltwo$).  The \hwv{} of each module is bosonic with $h_0$-charge $\lambda$ and conformal dimension $\frac{\lambda (\lambda+1)}{2(2\kk+3)}$.
	\item The ${}^{\NS} \aospdisp{\lambda}$ (${}^{\R} \aospdisp{\lambda}$), where $\lambda \notin \ZZ_{\ge 0}$.  Each is an irreducible \hwm{} in the \ns{} (Ramond) sector whose space of ground states forms an irreducible infinite-dimensional Verma module for $\osp$ ($\sltwo$).  The \hwv{} of each module is likewise bosonic with $h_0$-charge $\lambda$ and conformal dimension $\frac{\lambda (\lambda+1)}{2(2\kk+3)}$.
	\item The ${}^{\NS} \aospdism{\lambda}$ (${}^{\R} \aospdism{\lambda}$), where $\lambda \notin \ZZ_{\ge 0}$, that are the conjugates of the ${}^{\NS} \aospdisp{\lambda}$ (${}^{\R} \aospdisp{\lambda}$).  Conjugation for $\aosp$-modules also corresponds to twisting by the Weyl reflection of $\sltwo \ira \osp$.
	\item The ${}^{\NS} \aosprel{\lambda}{\Sigma}$ (${}^{\R} \aosprel{\lambda}{q}$), where $\lambda \in \CC$ and $\Sigma \in \CC$ ($q \in \CC$) satisfy $\lambda \neq \pm(\Sigma-\frac{1}{2}) \pmod{2}$ ($\lambda \neq -1 \pm \sqrt{1+2q} \pmod{2}$).  Each is an irreducible relaxed \hwm{} whose ground states have $h_0$-charges equal to $\lambda \pmod{2}$ and conformal dimension given by
	\begin{equation} \label{eq:osprelconfdims}
		{}^{\NS} \aosprel{\lambda}{\Sigma} : \quad \frac{\Sigma^2 - 1/4}{2(2\kk+3)}, \qquad
		{}^{\R} \aosprel{\lambda}{q} : \quad \frac{q - \kk/4}{2\kk+3}.
	\end{equation}
	The ground state of $h_0$-charge $\lambda$ is bosonic.  Here, $\Sigma$ denotes the eigenvalue of the $\osp$ super-Casimir \cite{ArnCas97}
	\begin{equation} \label{eq:defscas}
		\scas = x_0 y_0 - y_0 x_0 + \frac{1}{2}
	\end{equation}
	on the bosonic ground states, while $q$ denotes the ground state eigenvalue of the $\sltwo$ quadratic Casimir
	\begin{equation} \label{eq:defslcas}
		\cas = \frac{1}{2} h_0^2 + e_0 f_0 + f_0 e_0.
	\end{equation}
	\item The parity reversals of the above irreducibles obtained by declaring in each case that the ground state of $h_0$-charge $\lambda$ is fermionic rather than bosonic.  Parity reversal will be denoted by $\parrev$.
\end{itemize}
Of course, the spectral flows of these irreducible relaxed \hwms{} will again be irreducible, though they are usually not relaxed nor \hw{}.

We remark that this classification of irreducible relaxed \hw{} $\aosp$-modules relies crucially on the classification of irreducible weight $\osp$-modules with finite-dimensional weight spaces.  The latter result seems to have first appeared as part of the classification of \emph{all} irreducible weight $\osp$-modules, due to Bavula and van~Oystaeyen \cite{BavSim00}.  An elementary proof treating only the case of finite-dimensional weight spaces may be found in \cite[Thm.~2]{RidAdm17}.

\subsection{The coset construction} \label{sec:coset}

It is well known, see \cite{FanMod93} for an early instance and \cite[Thm.~8.4]{CreCos14} for a proof, that the coset (commutant) of an $\osp$ minimal model by its $\sltwo$ minimal model bosonic subalgebra (of the same level $\kk$) is a Virasoro minimal model.  Equating the expressions for $\kk$ from \eqref{eq:admksl2} and \eqref{eq:admkosp} gives
\begin{equation}
\kk+2=\frac{u}{v} \quad \text{and} \quad \kk+\frac 3 2=\frac{p}{2v}, \quad\text{where}\quad p+v=2u.
\end{equation}
The coset is then as in \eqref{coset}:
\begin{equation}
\vmodel{p,u} \cong \commutant{\slmodel{u,v}}{\ospmodel{p,v}} \equiv \frac{\ospmodel{p,v}}{\slmodel{u,v}}.
\end{equation}
Note that if $\ospmodel{p,v}$ is unitary, then both $\vmodel{p,u}$ and $\slmodel{u,v}$ must be unitary.  Thus, we must have $p-u=\pm1$ and $v=1$.  The only solution is $p=3$, $u=2$ and $v=1$, hence the only unitary $\osp$ minimal model is $\ospmodel{3,1}$ corresponding to $\kk = 0$ (this is the trivial one-dimensional \svoa{}).

In the remainder of the section, we shall discuss a proof of the coset identification \eqref{coset}.  The only step which we omit is that which establishes a particular character identity, \eqref{eq:charid} below, whose somewhat lengthy proof has already been detailed in \cite{CreRep17}.

At the level of the generating fields, the $\sltwo$ fields $e(z)$, $h(z)$ and $f(z)$ are identified with their namesakes in $\ospmodel{p,v}$, while the Virasoro field is identified with
\begin{equation}
T^{\supv}(z)=T^{\suposp}(z)-T^{\supsl}(z).
\end{equation}
This guarantees that $T^{\supv}$ has regular \opes{} with $e$, $h$ and $f$ \cite{GodVir85}.  Let $\VOA{V}_{\kk}$ denote the tensor product of the universal Virasoro \voa{} of central charge $1 - \frac{6(p-u)^2}{pu}$ and the universal $\sltwo$ \voa{} of level $\kk$.  The field identifications above then define a homomorphism of $\VOA{V}_{\kk}$ into $\ospmodel{p,v}$.

To show that this descends to an embedding
\begin{equation} \label{eq:embedding}
	\vmodel{p,u} \otimes \slmodel{u,v} \lira \ospmodel{p,v}
\end{equation}
and prove \eqref{coset}, we claim that it suffices to prove the following branching rule:
\begin{equation}
\res{\ospmodel{p,v}} \cong \bigoplus_{i=1}^{u-1} \virr{1,i} \otimes \aslfin{i,0}.\label{voabr}
\end{equation}
Here, we decompose $\ospmodel{p,v}$ as a $\VOA{V}_{\kk}$-module and note that the direct summands which appear are in fact $\vmodel{p,u} \otimes \slmodel{u,v}$-modules.  The embedding \eqref{eq:embedding} is now clear and the commutant of $\slmodel{u,v}$, here identified with its vacuum module $\aslfin{1,0}$, is obviously $\virr{1,1}$, the vacuum module of $\vmodel{p,u}$, as claimed.

As $\vmodel{p,u}$ is rational \cite{WanRat93,RidJac14} and $\slmodel{u,v}$ is rational in category $\categ{O}$ \cite{AdaVer95,DonVer97}, \eqref{voabr} will be proven if we can demonstrate its character analogue:
\begin{equation} \label{eq:charid}
	\fch{\ospmodel{p,v}} = \traceover{\ospmodel{p,v}} \zz^{h_0} \qq^{L^{\suposp}_0 - \cc^{\suposp}/24} = \sum_{i=1}^{u-1} \fchvir{1,i}{p,u} \fch{\aslfin{i,0}}.
\end{equation}
This is a straightforward, though somewhat lengthy, computation and is detailed in \cite[Lem.~2.1]{CreRep17}.  Actually, this calculation is performed at the level of meromorphic continuations of characters in $z \in \CC$ and $\abs{q}<1$, rather than as formal power series, hence its validity also requires the linear independence of these continuations (or careful attention to convergence regions).  Unfortunately, the continuations of the irreducible $\slmodel{u,v}$-characters in category $\categ{O}$ are not linearly independent if $v>1$ \cite{RidSL208}.  We can rectify this by replacing category $\categ{O}$ by its Kazhdan-Lusztig (or ordinary) subcategory $\klcat$ whose objects are the $\slmodel{u,v}$-modules in $\categ{O}$ with finite-dimensional $L^{\supsl}_0$-eigenspaces.  The irreducible characters in $\klcat$, which are precisely those of the $\aslfin{i,0}$, have linearly independent meromorphic continuations and so the above manipulations are justified and the proof is complete.

\subsection{Vertex tensor categories} \label{sec:vtcs}

The theory of vertex algebra extensions allows one to analyse vertex algebra constructions, such as the coset construction, in a purely categorical way.  This is based on the result that commutative and associative algebras in a given vertex tensor category are the same as vertex algebra extensions (in this category) \cite{HuaBra15}. In the case of vertex operator superalgebras, one has to work with commutative and associative superalgebras \cite{CreSim15}. We will not give precise definitions of the categorical terms here, instead referring to \cite{EtiTen15,CreTen17} for details.

In this section, we summarise the results of \cite{CreTen17} that are needed in what follows. The main result of that article is that the category of extended vertex superalgebra modules is braided-equivalent to the category of local modules for the corresponding algebra object. Moreover, there is an induction functor from the base category and this functor is braided-tensor, meaning in particular that it preserves the fusion rules.

The setup is as follows. Let $\VOA{V}$ be a simple \voa{} with integer conformal weights and let $\VOA{W}$ be a simple \svoa{}.  Assume that we have a parity-preserving embedding $\VOA{V} \ira \VOA{W}$, meaning that the image is contained in the bosonic subalgebra of $\VOA{W}$. This means that $\VOA{W}$ is an extension of $\VOA{V}$ and so it decomposes into $\VOA{V}$-modules as
\begin{equation} \label{eq:Wdecomp}
		\res{\VOA{W}} \cong \bigoplus_i \Mod{W}_i.
\end{equation}
Here and below, we assume that each of the $\Mod{W}_i$ consists of either bosonic or fermionic states.  An especially nice situation is when the $\Mod{W}_i$ appearing in this decomposition are irreducible and inequivalent. The notion $\res{\VOA{W}}$ of the restriction of $\VOA{W}$ to a module of the smaller \svoa{} $\VOA{V}$ generalises to arbitrary $\VOA{W}$-modules $\Mod{N}$ as we may also restrict them to $\VOA{V}$-modules:
\begin{equation}
		\res{\Mod{N}} \cong \bigoplus_j \Mod{N}_j.
\end{equation}
The identification of a restricted $\VOA{W}$-module, as a $\VOA{V}$-module, is called a \emph{branching rule}.

In this setup, there is a very closely related operation on modules called induction. For this, let $\Mod{M}$ be a $\VOA{V}$-module and consider its fusion product with the $\VOA{V}$-module $\res{\VOA{W}}$.  In many cases, the result has a natural structure as a $\VOA{W}$-module and this $\VOA{W}$-module is called the induction of $\Mod{M}$, denoted by $\ind{\Mod{M}}$. The restriction of an induced module decomposes as
\begin{equation} \label{eq:resind}
	\resind{\Mod{M}} \cong \bigoplus_i \Mod{W}_i \fuse \Mod{M}.
\end{equation}
Not every module induces to a local (meaning \ns{}) or twisted (meaning Ramond) module of $\VOA{W}$. Fortunately, there is a nice criterion to study the result of inducing, assuming that the conformal dimensions of the states of $\VOA{W}$ are integers (which is the case we are interested in here).  This criterion says that an induced module is local if and only if the twist acts as a $\VOA{W}$-module morphism. The twist is given by the action of $e^{2\pi i L_0}$, where $L_0$ is the Virasoro zero mode of $\VOA{W}$.  It follows that an irreducible induced module is \ns{} if and only if its conformal dimensions all differ by integers.  Moreover, an irreducible induced module is Ramond if and only if the conformal dimensions of its bosonic states differ from those of its fermionic ones by $\frac{1}{2}$ modulo $\ZZ$.

Let now $\Mod{M}_1$ and $\Mod{M}_2$ be two irreducible $\VOA{V}$-modules that both induce to irreducible $\VOA{W}$-modules. We ask the question of whether their inductions are isomorphic or not.  For this, there is a useful criterion called \emph{Frobenius reciprocity}. For our purposes, we may take it to say that the space of homomorphisms between two induced modules may be computed as
\begin{equation}\label{eq:frob}
	\hom_{\VOA{W}}(\ind{\Mod{M}_1}, \ind{\Mod{M}_2})\cong \hom_{\VOA{V}}({\Mod{M}_1}, \resind{\Mod{M}_2})\cong \bigoplus_i\hom_{\VOA{V}}({\Mod{M}_1}, \Mod{W}_i \fuse \Mod{M}_2).
\end{equation}
One therefore needs only to verify if $\Mod{M}_1$ appears as a submodule of the fusion product of any of the $\Mod{W}_i$ with $\Mod{M}_2$. We however warn the reader that one has to be careful with parity: in this setup, Frobenius reciprocity does not distinguish modules from their parity reversals.

We now come to the two most important statements of \cite{CreTen17}; we formulate them as theorems. The first one gives a criterion that guarantees that induced modules are irreducible.  We shall apply it frequently in what follows.
\begin{theorem}[\protect{\cite[Prop.~4.4]{CreTen17}}] \label{thm:indsimple}
	Let $\VOA{V} \ira \VOA{W}$ be an embedding of a simple \voa{} $\VOA{V}$ into a simple \svoa{} $\VOA{W}$  under which $\res{\VOA{W}}$ decomposes into a direct sum of irreducible $\VOA{V}$-modules $\Mod{W}_i$ as in \eqref{eq:Wdecomp}.  Suppose that $\Mod{M}$ is an irreducible $\VOA{V}$-module for which the fusion products $\Mod{W}_i \fuse \Mod{M}$ are irreducible and inequivalent: $\Mod{W}_i \fuse \Mod{M} \ncong \Mod{W}_j \fuse \Mod{M}$ if $i \neq j$.  Then, the induced $\VOA{W}$-module $\ind{\Mod{M}} = \VOA{W} \fuse \Mod{M}$ is irreducible.
\end{theorem}
\noindent Obviously, a necessary condition for the inequivalence of the $\Mod{W}_i \fuse \Mod{M}$ is that the $\Mod{W}_i$ are all inequivalent.

The second \lcnamecref{thm:induction} gives a way to easily determine the fusion rules of induced modules.  In categorical language, it states that induction is a vertex tensor functor.  The version below, which suffices for the application to follows, eschews this language for simplicity.
\begin{theorem}[\protect{\cite[Thm.~3.68]{CreTen17}}] \label{thm:induction}
	Let $\VOA{V} \ira \VOA{W}$ be an embedding of a \voa{} $\VOA{V}$ into an \svoa{} $\VOA{W}$ and let $\Mod{M}$ and $\Mod{N}$ be $\VOA{V}$-modules.  Then, the fusion rules of the induced $\VOA{W}$-modules satisfy
	\begin{equation} \label{eq:indfusion}
		\ind{\Mod{M}} \fuse \ind{\Mod{N}} \cong \ind{(\Mod{M} \fuse \Mod{N})}.
	\end{equation}
\end{theorem}
\noindent This method for computing fusion rules from \eqref{eq:indfusion} has also been proposed in the physics literature, for example in \cite[Eq.~(3.3)]{RidVer14}.


\section{Inverting the coset} \label{sec:construct}

Recall that the restriction of $\ospmodel{p,v}$ to an $\vmodel{p,u}\otimes\slmodel{u,v}$-module decomposes as in \eqref{voabr}.  The opposite operation, the induction of an $\vmodel{p,u}\otimes\slmodel{u,v}$-module $\Mod{M}$ to a $\ospmodel{p,v}$-module $\ind{\Mod{M}}$, is then defined by
\begin{equation}
	\ind{\Mod{M}} = \ospmodel{p,v} \fuse \Mod{M} \quad \Ra \quad
	\resind{\Mod{M}} \cong \bigoplus_{i=1}^{u-1} (\virr{1,i} \otimes \aslfin{i,0}) \fuse \Mod{M}, \label{indM2}
\end{equation}
where $\fuse$ denotes the fusion product of $\vmodel{p,u}\otimes\slmodel{u,v}$-modules.  In this section, we shall use induction to construct $\ospmodel{p,v}$-modules from $\vmodel{p,u}\otimes\slmodel{u,v}$-modules and identify them as level-$\kk$ $\aosp$-modules.  This is an instance of what we call ``inverting the coset''.  In \cref{sec:complete}, we will show that, up to isomorphism, every irreducible relaxed \hw{} $\ospmodel{p,v}$-module may be obtained in this fashion.

We start by recalling the branching rule \eqref{voabr}, in which $\ospmodel{p,v}$ is decomposed into irreducible $\vmodel{p,u}\otimes\slmodel{u,v}$-modules, and exploring the results of inducing its direct summands $\virr{1,i} \otimes \aslfin{i,0}$.  If $i=1$, then it is straightforward to identify the result, \emph{as an $\vmodel{p,u}\otimes\slmodel{u,v}$-module}, using the fusion rules \eqref{eq:frvir} and \eqref{eq:frsl2Lx}:
\begin{align}
	\resind{(\virr{1,1} \otimes \aslfin{1,0})} &\cong \bigoplus_{i=1}^{u-1} (\virr{1,i} \otimes \aslfin{i,0}) \fuse (\virr{1,1} \otimes \aslfin{1,0})
	\cong \bigoplus_{i=1}^{u-1} (\virr{1,i} \fuse \virr{1,1}) \otimes (\aslfin{i,0} \fuse \aslfin{1,0}) \notag \\
	&\cong \bigoplus_{i=1}^{u-1} \virr{1,i} \otimes \aslfin{i,0} \cong \res{\ospmodel{p,v}}.
\end{align}
The result, which is also obtained if $i=u-1$, is consistent with $\ind{(\virr{1,1} \otimes \aslfin{1,0})} \cong \ospmodel{p,v}$.  However, this does not by itself allow us to conclude that we have the corresponding isomorphism of $\ospmodel{p,v}$-modules.

Of course, $\ind{(\virr{1,1} \otimes \aslfin{1,0})} \cong \ospmodel{p,v}$ follows immediately from the definition of induction because $\virr{1,1} \otimes \aslfin{1,0}$ is just the vacuum module for $\vmodel{p,u} \otimes \slmodel{u,v}$.  However, this issue with identifying inductions is less trivial for other modules.  We shall therefore analyse this simple case in detail, describing a methodology that generalises straightforwardly to all modules.

Before commencing this analysis, we note that the induction is quite different for all $i \neq 1, u-1$.  For example, when $u>3$, we have
\begin{equation}
	\resind{(\virr{1,2} \otimes \aslfin{2,0})}
	\cong \bigoplus_{i=1}^{u-1} \virr{1,i} \otimes \aslfin{i,0} \oplus \bigoplus_{i=2}^{u-2} (\virr{1,i} \otimes \aslfin{i,0} \oplus \virr{1,i-1} \otimes \aslfin{i+1,0} \oplus \virr{1,i+1} \otimes \aslfin{i-1,0})
	\cong \res{\ospmodel{p,v}} \oplus \res{\Mod{M}},
\end{equation}
where $\Mod{M}$ is some other, as yet uncharacterised, $\ospmodel{p,v}$-module.  These results are consistent with \cref{thm:indsimple} which applies when $i$ is such that the $(\virr{1,i} \otimes \aslfin{i,0}) \fuse (\virr{1,j} \otimes \aslfin{j,0})$ are inequivalent and irreducible for all $j$.  If this holds, then the result of inducing is an irreducible $\ospmodel{p,v}$-module (which is clearly not the case in the previous example).

\subsection{The $\osp$ minimal models $\ospmodel{p,1}$} \label{subsec:unitary}

We start with the non-negative integer-level models $\ospmodel{p,1}$.  Here, $p$ is odd and greater than $1$, so $u = \kk+2 = \frac{p+1}{2} \ge 2$.  For these models, the only irreducible modules available for induction are the $\virr{r,s} \otimes \aslfin{r',0}$, where $r=1,\dots,p-1$ and $r',s=1,\dots,u-1$.  Inspecting the fusion rules involving these irreducibles and the $\virr{1,i} \otimes \aslfin{i,0}$, using \eqref{eq:frvir} and \eqref{eq:ufrsl2}, it is easy to see that the result will be irreducible if $r',s \in \set{1,u-1}$.  In this \lcnamecref{subsec:unitary}, we shall first analyse the intricacies regarding the induction procedure in some detail.  We then explain how to identify the induced module, including its sector (\ns{} or Ramond), given the parameters $r$, $s$ and $r'$.

Taking $r'=s=1$, we detail the determination of the decomposition of the induced module $\ind{(\virr{r,1} \otimes \aslfin{1,0})}$, which we shall denote by $\aospfin{r,0}$ for brevity:
\begin{equation} \label{unibr}
	\res{\aospfin{r,0}} \cong \bigoplus_{i=1}^{u-1} (\virr{1,i} \otimes \aslfin{i,0}) \fuse (\virr{r,1} \otimes \aslfin{1,0}) \cong \bigoplus_{i=1}^{u-1} \virr{r,i} \otimes \aslfin{i,0}.
\end{equation}
The summands on the \rhs{} are clearly inequivalent (and irreducible), hence \cref{thm:indsimple} applies and we conclude that $\aospfin{r,0}$ is an irreducible $\ospmodel{p,1}$-module as claimed.  However, taking $r'=u-1$ and $s=1$, $r'=1$ and $s=u-1$, or $r'=s=u-1$ gives inductions whose decompositions are identical to that in \eqref{unibr}, though perhaps with $r$ replaced by $p-r$.  For example, writing $\aospfintemp{r,0}$ for the irreducible $\ospmodel{p,1}$-module $\ind{(\virr{r,1} \otimes \aslfin{u-1,0})}$, we have
\begin{equation}\label{eqn:SameDecompSameModule}
	\res{\aospfintemp{p-r,0}} \cong \bigoplus_{i=1}^{u-1} \virr{p-r,i} \otimes \aslfin{u-i,0} = \bigoplus_{i=1}^{u-1} \virr{p-r,u-i} \otimes \aslfin{i,0} = \bigoplus_{i=1}^{u-1} \virr{r,i} \otimes \aslfin{i,0}.
\end{equation}
As before, however, this need not imply that $\aospfin{r,0}$ and $\aospfintemp{p-r,0}$ are isomorphic as $\ospmodel{p,1}$-modules.  As we shall see, they need not be.

To answer this question of possible isomorphisms, and to identify the induced modules $\aospfin{r,0}$ as $\aosp$-modules, we present two approaches.  The first method uses Frobenius reciprocity \eqref{eq:frob}.  Start by noting that the isomorphism \eqref{eqn:SameDecompSameModule} gives an inclusion of $\virr{r,1}\otimes\aslfin{1,0}$ into $\res{\aospfintemp{p-r,0}}$.  Reciprocity then says that there is a non-zero map from $\aospfin{r,0} = \ind{(\virr{r,1}\otimes\aslfin{1,0})}$ to $\aospfintemp{p-r,0}$.  As both modules are known to be irreducible, this map is an isomorphism by Schur's lemma.

But, we have stated that these modules need not be isomorphic!  The problem arises because we really want to determine if there is an isomorphism between $\aospfin{r,0}$ and $\aospfintemp{p-r,0}$ that preserves the parity of the vectors (as we distinguish between a $\ospmodel{p,1}$-module and its parity reversal).  Our application of Frobenius reciprocity started with the existence of a map between $\vmodel{p,u} \otimes \slmodel{u,1}$-modules, which have no concept of parity, it follows that the deduced map between $\ospmodel{p,1}$-modules need not respect parity.  More precisely, it could respect or reverse parity.

To ameliorate this deficiency, we describe a second, more instructive, approach which relies on explicitly identifying the ground states of the (irreducible) induced module.  This has the added advantage of allowing us to compare with the list of irreducible $\aosp$-modules given in \cref{sec:osp} and thereby identify the induced module completely.

The ground states of the irreducible induced module $\aospfin{r,0}$ are easily found by determining which of the ground states of the summands $\virr{r,i} \otimes \aslfin{i,0}$ appearing in \eqref{unibr} have the lowest conformal dimension.  By \eqref{eq:vconfdim} and \eqref{eq:usl2weights}, the conformal dimension of the ground states of the $i$-th summand is
\begin{equation}
	\Delta^{\supv}_{r,i} + \Delta^{\supsl}_{i,0} = \frac{1}{2} i^2 - \frac{r}{2} i + \frac{(r^2-1)u}{4p}.
\end{equation}
The global minimum therefore occurs when $i=\frac{1}{2} r$, if $r$ is even, and when $i=\frac{1}{2} (r\pm1)$, if $r$ is odd.  This minimal conformal dimension may now be written in the form
\begin{equation}
\Delta^{\suposp}_{r,0}=\frac{r^2-1}{8p}-\frac{1+(-1)^r}{16}.\label{unicondim}
\end{equation}
Moreover, the ground states of minimal conformal dimension have a \hwv{} whose $h_0$-charge is
\begin{equation}
\lambda^{\suposp}_{r,0}=\frac{r-1}{2}-\frac{1+(-1)^{r}}{4}.\label{unicharge}
\end{equation}
$\aospfin{r,0}$ is therefore an irreducible \hw{} $\aosp$-module of $h_0$-charge $\lambda^{\suposp}_{r,0}$.  To determine its sector, note that the conformal dimensions of the ground states of the $i$-th and $j$-th summands in \eqref{unibr} differ by $\frac{1}{2} (i-j)(i+j-r)$.  If $r$ is odd, then this difference is always an integer so $\aospfin{r,0}$ belongs to the \ns{} sector.  Likewise, $\aospfin{r,0}$ belongs to the Ramond sector when $r$ is even.

It only remains to determine the parity of the \hwv{} of $\aospfin{r,0}$.  To do so, note that the $h_0$-charges of the summands $\virr{r,i} \otimes \aslfin{i,0}$ in \eqref{unibr} are equal to $i-1 \pmod{2}$.  As the states of $\virr{r,1} \otimes \aslfin{1,0}$ are bosonic (because this is the module we are inducing from), it follows that $\virr{r,i} \otimes \aslfin{i,0}$ is bosonic for $i$ odd and fermionic for $i$ even.  For $r$ odd, the \hwv{} corresponds to $i = \frac{r+1}{2}$, hence it is bosonic if $r=1 \pmod{4}$ and fermionic if $r=3 \pmod{4}$.  For $r$ even, we similarly conclude that we have a bosonic \hwv{} if $r=2 \pmod{4}$ and a fermionic one if $r=0 \pmod{4}$.  Comparing with the list of irreducible $\aosp$-modules given in \cref{sec:osp}, this then completes the identification of the $\aospfin{r,0}$.
\begin{center}
	\begin{tabular}{C|CCCC}
		r \pmod{4} & 1 & 2 & 3 & 4 \\
		\hline
		\aospfin{r,0} & {}^{\NS} \aospfin{\lambda^{\suposp}_{r,0}} & {}^{\R} \aospfin{\lambda^{\suposp}_{r,0}} & \parrev {}^{\NS} \aospfin{\lambda^{\suposp}_{r,0}} & \parrev ^{\R} \aospfin{\lambda^{\suposp}_{r,0}}
	\end{tabular}
\end{center}
We recall that $\parrev$ denotes parity reversal, meaning that the module has had its bosonic and fermionic subspaces swapped.  Note that $\ind{(\virr{1,1} \otimes \aslfin{1,0})} = \aospfin{1,0} \cong {}^{\NS} \aospfin{0}$ is indeed the vacuum module of $\ospmodel{p,1}$, as expected.

This table describes a dictionary that identifies the irreducible $\ospmodel{p,1}$-modules, obtained by induction, as irreducible level-$\kk$ \hw{} $\aosp$-modules, where $\kk = p-\frac{3}{2}$, with both parity and sector made explicit.  We shall find this dictionary, as well as those obtained in \cref{subsec:nonunimod} for the irreducible $\ospmodel{p,v}$-modules, extremely useful in what follows.

If we repeat this analysis with the $\aospfintemp{r,0}$, we do not obtain any new $\ospmodel{p,1}$-modules except perhaps for parity reversals.  Indeed, the identification is as follows.
\begin{center}
	{\renewcommand{\arraystretch}{1.1}
	\begin{tabular}{C|CCCC}
		r \pmod{4} & 1 & 2 & 3 & 4 \\
		\hline
		\aospfintemp{r,0} & {}^{\R} \aospfin{\lambda^{\suposp}_{p-r,0}} & {}^{\NS} \aospfin{\lambda^{\suposp}_{p-r,0}} & \parrev {}^{\R} \aospfin{\lambda^{\suposp}_{p-r,0}} & \parrev ^{\NS} \aospfin{\lambda^{\suposp}_{p-r,0}}
	\end{tabular}
	}
\end{center}
In particular, $\aospfin{r,0}$ is isomorphic to $\aospfintemp{p-r,0}$, if $p=3 \pmod{4}$, and to $\parrev \aospfintemp{p-r,0}$, if $p=1 \pmod{4}$.  We remark that the fact that no new modules are encountered (except parity reversals) was guaranteed because the spectral flow automorphisms of $\asltwo$ and $\aosp$ are consistent with the coset construction.  Thus,
\begin{equation}
	\aospfintemp{r,0} = \ind{(\virr{r,1} \otimes \aslfin{u-1,0})} \cong \ind{\brac*{\virr{r,1} \otimes \sfmodsl{\aslfin{1,0}}{}}} \cong \sfmodosp{\ind{(\virr{r,1} \otimes \aslfin{1,0})}}{} = \sfmodosp{\ind{\aospfin{r,0}}}{},
\end{equation}
a relation that is easy to verify directly.  We conclude that inducing the $\virr{r,1} \otimes \aslfin{1,0}$ and applying parity reversal will give all the irreducibles that can be obtained by inducing an arbitrary $\vmodel{p,u} \otimes \slmodel{u,v}$-module and parity-reversing.

The characters of the $\aospfin{r,0}$ are now obtained by taking characters of modules on both sides of the branching rule \eqref{unibr}.  This gives
\begin{equation}
\fch{\aospfin{r,0}}=\traceover{\aospfin{r,0}}\zz^{h_0}\qq^{L^{\suposp}_0-\cc^{\suposp}/24}=\sum_{i=1}^{u-1} \fchvir{r,i}{p,u}\, \fch{\aslfin{i,0}}.\label{unichar}
\end{equation}
One can expand this using the explicit forms \eqref{eq:chvir} and \eqref{eq:chsl2} for the irreducible $\vmodel{p,u}$- and $\slmodel{u,v}$-characters.
Since $\aosp$ is a superalgebra, it is appropriate to consider its supercharacters as well. As the \hwv{} of the module $\virr{r,1}\otimes\aslfin{1,0}$ is bosonic, its $h_0$-charge differs from those of the fermionic states by an odd integer. The supercharacter of $\aospfin{r,0}$ is therefore simply given by
\begin{equation} \label{unisupchar}
\fsch{\aospfin{r,0}}=\traceover{\aospfin{r,0}}(-1)^F \zz^{h_0}\qq^{L^{\suposp}_0-\cc^{\suposp}/24}
=\sum_{i=1}^{u-1} (-1)^{i-1} \fchvir{r,i}{p,u}\, \fch{\aslfin{i,0}},
\end{equation}
where $F$ acts as $0$ on a bosonic state and as multiplication by $1$ on a fermionic one.

\subsection{The $\osp$ minimal models $\ospmodel{p,v}$ with $v \neq 1$} \label{subsec:nonunimod}

Following a similar method as in the $v=1$ case, we construct irreducible $\ospmodel{p,v}$-modules from those of $\vmodel{p,u}$ and $\slmodel{u,v}$ through induction. These modules are then identified as $\aosp$-modules using the list presented in \cref{sec:osp}.  This identification uses $h_0$-charges and conformal dimensions and is therefore straightforward for all cases except that of the \ns{} relaxed \hwms{} ${}^{\NS} \aosprel{\lambda}{\Sigma}$ for which the super-Casimir eigenvalue $\Sigma$ on bosonic eigenstates is only determined by the conformal dimension up to a sign, see \eqref{eq:osprelconfdims}.

To fix this sign, we must realise $\Sigma$ in terms of $\vmodel{p,u}$ and $\slmodel{u,v}$ data.  Recall that the super-Casimir $\scas$, defined in \eqref{eq:defscas}, of $\osp$ (embedded in $\aosp$ as the horizontal subalgebra) commutes with $e_0$, $h_0$ and $f_0$, but anticommutes with $x_0$ and $y_0$.  We therefore introduce the field
\begin{equation}
\scas(z)=\normord{xy}(z)-\normord{yx}(z),
\end{equation}
noting that its zero mode $\scas_0$ acts on \ns{} ground states as multiplication by $\pm\Sigma - \frac{1}{2}$, where the sign is positive for bosonic ground states and negative for fermionic ones.  It is now straightforward to check that $\scas(z)$ is realised as
\begin{equation}
	\scas(z) = 2 T^{\supsl}(z) - \frac{2p}{v} T^{\supv}(z),
\end{equation}
under the embedding \eqref{eq:embedding}.  It follows that $\Sigma$ may be computed in terms of the action of the zero modes of $T^{\supsl}(z)$ and $T^{\supv}(z)$ acting on a bosonic \ns{} ground state $v$:
\begin{equation} \label{sCasAction}
	\Sigma v = \brac*{2L_0^{\supsl}-\frac{2p}{v} L_0^{\supv}+\frac{1}{2}} v.
\end{equation}

Having dealt with this minor subtlety, we can now follow the same procedure as in the $v=1$ case and construct irreducible $\ospmodel{p,v}$-modules by inducing certain modules of $\vmodel{p,u}\otimes\slmodel{u,v}$. We shall adopt the following convention in defining our $\ospmodel{p,v}$-modules:
\begin{gather}
\aospfin{r,0}= \ind{(\virr{r,1}\otimes\aslfin{1,0})},\quad
\aospdispm{r,s}=\ind{(\virr{r,1}\otimes\asldispm{1,s})},\quad
\aosprel{\lambda}{r,s}=\ind{(\virr{r,1}\otimes\aslrel{\lambda}{1,s})},\quad
\aosprelredpm{r,s}=\ind{(\virr{r,1}\otimes\aslrelredpm{1,s})}.\label{typesind}
\end{gather}
Here, $r=1,\dots,p-1$ and $s=1,\dots,v-1$, while $\lambda \in \CC$ satisfies $\lambda \neq \lambda^{\supsl}_{1,s}, \lambda^{\supsl}_{u-1,v-s} \pmod{2}$.  The corresponding branching rules are computed as in \eqref{unibr} and are given by
\begin{equation} \label{eq:genbr}
\res{\aospfin{r,0}} \cong \bigoplus_{i=1}^{u-1} \virr{r,i}\otimes\aslfin{i,0},\quad
\res{\aospdispm{r,s}} \cong \bigoplus_{i=1}^{u-1} \virr{r,i}\otimes\asldispm{i,s},\quad
\res{\aosprel{\lambda}{r,s}} \cong \bigoplus_{i=1}^{u-1} \virr{r,i}\otimes\aslrel{\lambda+i-1}{i,s},\quad
\res{\aosprelredpm{r,s}} \cong \bigoplus_{i=1}^{u-1} \virr{r,i}\otimes\aslrelredpm{i,s}.
\end{equation}
It is now easy to check that \cref{thm:indsimple} applies to the $\aospfin{r,0}$, $\aospdispm{r,s}$ and $\aosprel{\lambda}{r,s}$, hence that these are irreducible $\ospmodel{p,v}$-modules.

As before, the states in the $\vmodel{p,u}\otimes\slmodel{u,v}$-module being induced are bosonic in the resulting $\ospmodel{u,v}$-module, hence the states of the summands of \eqref{eq:genbr} with $i$ odd (even) are bosonic (fermionic). In each branching rule, we determine the indices $i$ for which the conformal dimension of the ground states of the $\vmodel{p,u}\otimes\slmodel{u,v}$-module is minimised.  In the \ns{} sector, where $r+s\in2\ZZ+1$, the global minimum occurs for $i = \frac{r+s\pm1}{2}$, while in the Ramond sector, where $r+s\in2\ZZ$, the minimum is at $i=\frac{r+s}{2}$.  (We take $s=0$ for the $\aospfin{r,0}$.)  The conformal dimensions of the ground states of the induced modules \eqref{typesind} are thereby found to be given by
\begin{equation} \label{condim}
  \Delta_{r,s}^{\suposp}=\frac{(vr-ps)^2-v^2}{8pv}-\frac{1+(-1)^{r+s}}{16}.
\end{equation}
This clearly reduces to \eqref{unicondim} when $v=1$ (forcing $s=0$).

The $\aospfin{r,0}$ and $\aospdisp{r,s}$ are \hw{} $\ospmodel{p,v}$-modules and the $h_0$-charges of their \hwvs{} are easily seen to be
\begin{equation}
	\lambda_{r,s}^{\suposp}=\frac{1}{2} \brac*{r-1-\frac{p}{v}s}-\frac{1+(-1)^{r+s}}{4}.
\end{equation}
This likewise reduces to \eqref{unicharge} when $v=1$ and $s=0$.  The $\aospdism{r,s}$ are clearly the conjugates of the $\aospdisp{r,s}$, so it remains to identify the $\aosprel{\lambda}{r,s}$ and the $\aosprelredpm{r,s}$.  In the \ns{} sector, we use \eqref{sCasAction} to show that the super-Casimir eigenvalue on the bosonic ground states is
\begin{equation} \label{eq:defSigma}
	\Sigma_{r,s} = \frac{1}{2} (-1)^{(r+s-1)/2} \brac*{r - \frac{p}{v} s},
\end{equation}
which is easily checked to be consistent with \eqref{eq:osprelconfdims} and \eqref{condim}.  In the Ramond sector, \eqref{eq:osprelconfdims} and \eqref{condim} lead to the eigenvalue of the $\sltwo$ Casimir on the ground states being
\begin{equation} \label{eq:defq}
	q_{r,s} = \frac{1}{8} \brac*{r - \frac{p}{v} s}^2 - \frac{1}{2}.
\end{equation}

We now summarise the properties of the induced $\ospmodel{p,v}$-modules \eqref{typesind} in the following list, thereby identifying them as $\aosp$-modules.  Modules with $r+s$ odd (even), where $s$ is understood to be $0$ for the $\aospfin{r,0}$, belong to the \ns{} (Ramond) sector.  The global parities of these induced modules are determined as in \cref{subsec:unitary}.
\begin{itemize}
	\item The $\aospfin{r,0}$, with $1\le r\le p-1$, are 	irreducible \hwms{} whose ground state spaces are finite-dimensional. The \hwv{} of each module has $h_0$-charge $\lambda^{\suposp}_{r,0}$ and conformal dimension $\Delta^{\suposp}_{r,0}$.  The sectors and global parities are found to follow the same pattern as for the case where $v=1$.
	\begin{center}
		\begin{tabular}{C|CCCC}
			r \pmod{4} & 1 & 2 & 3 & 4 \\
			\hline
			\aospfin{r,0} & {}^{\NS} \aospfin{\lambda^{\suposp}_{r,0}}
			              & {}^{\R} \aospfin{\lambda^{\suposp}_{r,0}}
			              & \parrev {}^{\NS} \aospfin{\lambda^{\suposp}_{r,0}}
			              & \parrev ^{\R} \aospfin{\lambda^{\suposp}_{r,0}}
		\end{tabular}
	\end{center}
	\item The $\aospdisp{r,s}$, with $1\le r\le p-1$ and $1\le s\le v-1$, are irreducible \hwms{} whose ground state spaces are infinite-dimensional. The \hwv{} has charge $\lambda^{\suposp}_{r,s}$ and conformal dimension $\Delta^{\suposp}_{r,s}$.  The $\aospdism{r,s}$ are the conjugates of the $\aospdisp{r,s}$.
	\begin{center}
		\begin{tabular}{C|CCCC}
			r+s \pmod{4} & 1 & 2 & 3 & 4 \\
			\hline
			\aospdispm{r,s} & {}^{\NS} \aospdispm{\lambda_{r,s}}
			                & {}^{\R} \aospdispm{\lambda_{r,s}}
			                & \parrev {}^{\NS} \aospdispm{\lambda_{r,s}}
			                & \parrev ^{\R} \aospdispm{\lambda_{r,s}}
		\end{tabular}
	\end{center}
	\item The $\aosprel{\lambda}{r,s}$, with $1\le r\le p-1$, $1\le s\le v-1$ and $\lambda \neq \lambda^{\supsl}_{1,s}, \lambda^{\supsl}_{u-1,v-s} \pmod{2}$ are irreducible relaxed \hwms{} whose ground state spaces are infinite-dimensional.  There is a bosonic ground state of charge $\lambda$ that is characterised by its super-Casimir eigenvalue $\Sigma_{r,s}$ (if $r+s$ is odd) or its $\sltwo$ Casimir eigenvalue $q_{r,s}$ (if $r+s$ is even).  In either case, the conformal dimension of the ground states is $\Delta^{\suposp}_{r,s}$.
	\begin{center}
		\begin{tabular}{C|CCCC}
			r+s \pmod{4} & 1 & 2 & 3 & 4 \\
			\hline
			\aosprel{\lambda}{r,s} & {}^{\NS} \aosprel{\lambda}{\Sigma_{r,s}}
			                       & {}^{\R} \aosprel{\lambda}{q_{r,s}}
			                       & {}^{\NS} \aosprel{\lambda}{\Sigma_{r,s}}
			                       & \parrev {}^{\R}\aosprel{\lambda+1}{q_{r,s}}
		\end{tabular}
	\end{center}
	It is easy to check that the restriction $\lambda \neq \lambda^{\supsl}_{1,s}, \lambda^{\supsl}_{u-1,v-s} \pmod{2}$ translates into $\lambda \neq \xi^{\pm}_{r,s} \pmod{2}$, where
	\begin{equation} \label{eq:translation}
		\xi^{\pm}_{r,s} =
		\begin{cases*}
			\pm (\Sigma_{r,s} - \frac{1}{2}), & if $r+s$ is odd, \\
			-1 \pm \sqrt{1+2q_{r,s}}, & if $r+s$ is even.
		\end{cases*}
	\end{equation}
	For example, $r+s=1 \pmod{4}$ implies that
	\begin{equation}
		\lambda^{\supsl}_{1,s} = -\frac{u}{v} s = \frac{1}{2} (-s - \frac{p}{v} s) = \frac{1}{2} (r-1 - \frac{p}{v} s) = \Sigma_{r,s} - \frac{1}{2} \pmod{2}
	\end{equation}
	and, similarly, $\lambda^{\supsl}_{u-1,v-s} = -(\Sigma_{r,s} - \frac{1}{2}) \pmod{2}$.
	\item The $\aosprelredpm{r,s}$, with $1\le r\le p-1$ and $1\le s\le v-1$, are reducible relaxed \hwms{} with a bosonic ground state of charge $\lambda^{\suposp}_{r,s}$ and conformal dimension $\Delta^{\suposp}_{r,s}$.  They are characterised by the following short exact sequences:
	\begin{equation} \label{eq:esosp}
		\dses{\aospdispm{r,s}}{\aosprelredpm{r,s}}{\parrev^u \aospdismp{p-r,v-s}}.
	\end{equation}
	Unpacking this, we find that the submodule $\Mod{S}$ and quotient $\Mod{Q}$ of $\aosprelredpm{r,s}$ are identified as follows.
	\begin{center}
		\begin{tabular}{C|CCCC}
			r+s \pmod{4} & 1 & 2 & 3 & 4 \\
			\hline
			\Mod{S} & {}^{\NS} \aospdispm{\lambda_{r,s}}
			        & {}^{\R} \aospdispm{\lambda_{r,s}}
			        & \parrev {}^{\NS} \aospdispm{\lambda_{r,s}}
			        & \parrev ^{\R} \aospdispm{\lambda_{r,s}} \\
			\Mod{Q} & \parrev {}^{\NS} \aospdismp{\lambda_{p-r,v-s}}
			        & {}^{\R} \aospdismp{\lambda_{p-r,v-s}}
			        & {}^{\NS} \aospdismp{\lambda_{p-r,v-s}}
			        & \parrev ^{\R} \aospdismp{\lambda_{p-r,v-s}}
		\end{tabular}
	\end{center}
\end{itemize}
We emphasise that the parity reversals of the $\aospfin{r,0}$, $\aospdispm{r,s}$, $\aosprel{\lambda}{r,s}$ and $\aosprelredpm{r,s}$ are also $\ospmodel{p,v}$-modules, as are their images under spectral flow.

The characters and supercharacters of the induced $\ospmodel{p,v}$-modules follow from \eqref{eq:genbr} as in the $v=1$ case. The characters are given by
\begin{subequations}
	\begin{align}
	\fch{\aospfin{r,0}} &= \sum_{i=1}^{u-1} \fchvir{r,i}{p,u}\, \fch{\aslfin{i,0}}, &
	\fch{\aospdispm{r,s}} &= \sum_{i=1}^{u-1} \fchvir{r,i}{p,u}\, \fch{\asldispm{i,s}}\\
	\fch{\aosprel{\lambda}{r,s}} &= \sum_{i=1}^{u-1} \fchvir{r,i}{p,u}\, \fch{\aslrel{\lambda+i-1}{\Delta_{i,s}}}, &
	\fch{\aosprelredpm{r,s}} &= \sum_{i=1}^{u-1}\fchvir{r,i}{p,u}\, \fch{\aslrelredpm{i,s}}
	\end{align}
\end{subequations}
and the supercharacters by the same formulae, but with $(-1)^{i-1}$ inserted into each sum.  More explicit formulae may now be obtained by substituting \eqref{eq:chvir} and \eqref{eq:chsl2}.  As usual, the characters and supercharacters of parity reversals are obtained from
\begin{equation}
	\ch{\parrev \Mod{M}} = \ch{\Mod{M}}, \qquad \sch{\parrev \Mod{M}} = -\sch{\Mod{M}}.
\end{equation}

We remark that substituting the formula \eqref{charE} for the irreducible relaxed $\slmodel{u,v}$-characters gives the following form for the irreducible relaxed $\ospmodel{p,v}$-characters:
\begin{equation}
	\fch{\aosprel{\lambda}{r,s}} = \frac{1}{\eta(\qq)^2} \sum_{i=1}^{u-1} \zz^{\lambda+i-1} \fchvir{r,i}{p,u}\,\fchvir{i,s}{u,v} \sum_{j\in\ZZ} \zz^{2j}.
\end{equation}
Comparing with the character formulae recently proved in \cite{KawRel18}, we deduce the following remarkable identities:
\begin{subequations} \label{eq:bilinvir}
	\begin{align}
		\sum_{i=1}^{u-1} \fchvir{r,i}{p,u} \, \fchvir{i,s}{u,v} &=
		\begin{dcases*}
			\fchsvir{r,s}{p,v} \sqrt{\frac{\fjth{3}{1;\qq}}{\eta(\qq)}}, & if $r+s \in 2\ZZ$, \\
			2 \fchsvir{r,s}{p,v} \sqrt{\frac{\fjth{2}{1;\qq}}{2 \eta(\qq)}}, & if $r+s \in 2\ZZ+1$,
		\end{dcases*} \label{eq:chbilinvir}
		\\
		\sum_{i=1}^{u-1} (-1)^{i-1} \fchvir{r,i}{p,u} \, \fchvir{i,s}{u,v} &=
		\begin{dcases*}
			\fschsvir{r,s}{p,v} \sqrt{\frac{\fjth{4}{1;\qq}}{\eta(\qq)}}, & if $r+s \in 2\ZZ$, \\
			0, & if $r+s \in 2\ZZ+1$.
		\end{dcases*}\label{eq:schbilinvir}
	\end{align}
\end{subequations}
Here, $\chsvir{r,s}{p,v}$ and $\schsvir{r,s}{p,v}$ denote the characters and supercharacters of the $N=1$ superconformal minimal model $\svmodel{p,v}$ of central charge $\frac{3}{2} - \frac{3(v-p)^2}{pv}$:
\begin{subequations}
	\begin{align}
		\fchsvir{r,s}{p,v} &=
		\begin{dcases*}
			\frac{1}{\eta(\qq)} \sqrt{\frac{\fjth{3}{1;\qq}}{\eta(\qq)}} \sum_{n \in \ZZ} \sqbrac*{\qq^{(2npv+vr-us)^2/8pv} - \qq^{(2npv+vr+us)^2/8pv}}, & if $r+s \in 2\ZZ$, \\
			\frac{1}{\eta(\qq)} \sqrt{\frac{\fjth{2}{1;\qq}}{2 \eta(\qq)}} \sum_{n \in \ZZ} \sqbrac*{\qq^{(2npv+vr-us)^2/8pv} - \qq^{(2npv+vr+us)^2/8pv}}, & if $r+s \in 2\ZZ+1$,
		\end{dcases*}
		\\
		\fschsvir{r,s}{p,v} &=
		\begin{dcases*}
			\frac{1}{\eta(\qq)} \sqrt{\frac{\fjth{4}{1;\qq}}{\eta(\qq)}} \sum_{n \in \ZZ} (-1)^{np} \sqbrac*{\qq^{(2npv+vr-us)^2/8pv} - (-1)^r \qq^{(2npv+vr+us)^2/8pv}}, & if $r+s \in 2\ZZ$, \\
			0, & if $r+s \in 2\ZZ+1$.
		\end{dcases*}
	\end{align}
\end{subequations}
The identities \eqref{eq:bilinvir} may be understood as resulting from the branching rules for the coset described by the embedding
\begin{equation} \label{eq:NiceCoset}
	\vmodel{p,u} \otimes \vmodel{u,v} \lira \svmodel{p,v} \otimes \VOA{F},
\end{equation}
where $\VOA{F}$ denotes the free fermion \svoa{}.  Indeed, this is strongly suggested by the character decomposition \eqref{eq:chbilinvir} with $r=s=1$ and is easily confirmed by explicitly constructing the two commuting Virasoro subalgebras.  A version of this coset was previously considered, but deduced heuristically, in \cite{CrnRen89,LasSup93} --- however, there $\VOA{F}$ was incorrectly replaced by its bosonic orbifold $\vmodel{3,4}$.  From our perspective, it is natural to regard this beautiful coset as the quantum hamiltonian reduction of the coset \eqref{coset} (this is explained in \cite[Thm.~2.10]{CrS18} and \cite{CKM2}).

\section{Completeness of the irreducible spectrum} \label{sec:complete}

In the previous \lcnamecref{sec:construct}, we have constructed several families of irreducible $\ospmodel{p,v}$-modules using $\vmodel{p,u}$- and $\slmodel{u,v}$-modules as building blocks.  A natural question to ask is whether this procedure has in fact constructed \emph{all} the irreducible $\ospmodel{p,v}$-modules, up to isomorphism.  The answer to this is surely no, because one expects to be able to similarly construct irreducible Whittaker modules for $\ospmodel{p,v}$ from those known for $\slmodel{u,v}$ when $v>1$ \cite{AdaRea17}.  However, we can refine our question to instead ask whether we have constructed all the irreducible $\ospmodel{p,v}$-modules in some physically relevant, and hopefully consistent, class (category) of $\aosp$-modules.

When $v=1$, this question was asked and answered in \cite{CreRep17} using the notion of Perron-Frobenius dimensions for the $\aosp$-analogue of the Kazhdan-Lusztig category $\klcat$ discussed at the end of \cref{sec:coset}.  This relied crucially on there being only finitely many irreducible \hw{} $\ospmodel{p,v}$-modules, up to isomorphism.  As such, this dimension argument should also succeed when $v>1$ as long as we only want to know if we have constructed all the irreducible \hw{} $\ospmodel{p,v}$-modules with finite-dimensional $L^{\suposp}_0$-eigenspaces.  It will not obviously help with the completeness question for more general classes of modules.

Here, we shall instead use a different tool, Zhu's algebras, to prove that the lists of irreducible relaxed \hw{} $\ospmodel{p,v}$-modules constructed in \cref{sec:construct} are complete.  We strongly believe that there is a physically consistent category for these \svoas{} in which the simple objects are precisely the spectral flows of the irreducible relaxed \hwms{}.  It therefore suffices to complete the classification of irreducible relaxed \hw{} $\ospmodel{p,v}$-modules.  We shall first do this for the case $v=1$, for which there is an easy argument, independent of our constructions, that trivially recovers the classification result of \cite{CreRep17}.  We shall then present a slightly more involved argument for $v>1$ that relies on our constructions to provide a quick proof of the general classification.  This classification was originally proved in \cite[Thm.~3.7]{WooAdm18} using symmetric functions.

\subsection{Zhu's algebra} \label{sec:Zhu}

There are two basic observations, both very familiar to physicists, that underlie the formalism developed by Zhu \cite{ZhuMod96}, see also \cite{FreVer92,KacVer94,DonTwi98}, to classify suitably nice \svoa{} modules.  The first is that for irreducible relaxed \hw{} modules (see \cite{RidRel15} for a general definition), one can completely identify the module from the action of the zero modes of the algebra on the ground states (relaxed \hwvs{}) of the module.  Here, the zero mode of the field $v(z)$, assumed to have definite conformal dimension $\Delta$, is given by
\begin{equation}
	v_0 = \oint_0 v(z) z^{\Delta-1} \, \frac{\dd z}{2 \pi \ii},
\end{equation}
as usual.  The second observation is that ``setting singular vectors to zero'' in the vacuum module, hence in the \svoa{}, results in zero modes that must annihilate the ground states, thereby constraining the representation theory.  This was implicitly used in Gepner and Witten's analysis \cite{GepStr86} of WZW models and was made explicit in Feigin, Nakanishi and Ooguri's work on Virasoro minimal models \cite{FeiAnn92}.

Zhu's algebra is thus nothing more than the algebra of zero modes of the \svoa{}, constrained to act on ground states.  In fact, there are two Zhu algebras in the super-setting, one for \ns{} ground states and another for Ramond ones (the difference lies in which fields actually have zero modes).  The first basic observation above is now formalised as the following beautiful correspondence between \svoa{} modules and Zhu algebra modules.
\begin{theorem}[\cite{ZhuMod96}] \label{thm:zhu}
	Let $\VOA{V}$ be a \svoa{} and let $\zhu{}{\VOA{V}}$ be its Zhu algebra (\ns{} or Ramond).  Then:
	\begin{enumerate}
		\item The ground states of an irreducible relaxed \hw{} $\VOA{V}$-module naturally form an irreducible weight $\zhu{}{\VOA{V}}$-module.
		\item An irreducible weight $\zhu{}{\VOA{V}}$-module may always be induced to an irreducible relaxed \hw{} $\VOA{V}$-module whose ground states realise the original $\zhu{}{\VOA{V}}$-module.
		\item These correspondences give rise to a bijection between the isomorphism classes of irreducible relaxed \hw{} $\VOA{V}$-modules and irreducible weight $\zhu{}{\VOA{V}}$-modules.
	\end{enumerate}
\end{theorem}
\noindent This shifts the question of classifying irreducible relaxed \hwms{} for a \svoa{} to the (hopefully easier) question of classifying irreducible weight modules for an associative algebra.

The formal definitions \cite{ZhuMod96,DonTwi98} of the Zhu algebras are, unfortunately, usually given in a form which obfuscates this simple origin.  We shall therefore not discuss these general definitions, but instead use the equivalent, but more practical, definition described in \cite[Sec.~4.1]{RidAdm17} (for example) for affine \svoas{}.  The equivalence of the zero mode and formal definitions is discussed, in varying degrees of detail, in \cite[Lect.~18.5--6]{KacBom13}, \cite[App.~B]{RidRel15} and \cite[App.~A]{BloSVir16}.

Let $\VOA{V}_{\kk}$ denote the universal level-$\kk$ \svoa{} associated to $\osp$, where $\kk \neq -\frac{3}{2}$ is non-critical.  Let $\aoenvalgk$ denote the quotient of the \uea{} of $\aosp$ by the ideal generated by $K - \kk \wun$ and let $\aoenvalgk^0$ be its conformal weight zero subalgebra (the centraliser of $L_0$ in $\aoenvalgk$).  Then, there is a projection $\pi_0$ from $\aoenvalgk^0$ into $\oenvalg$, the \uea{} of $\osp$, whose kernel is spanned by the Poincar\'{e}-Birkhoff-Witt (PBW) basis elements, ordered by increasing mode index, that involve at least one mode with a non-zero index.  (Here, we are identifying zero modes with elements of $\osp$.)  The \ns{} Zhu algebra $\nszhu{\VOA{V}_{\kk}}$ is then the image of the map $v \in \VOA{V}_{\kk} \mapsto [v] = \pi_0(v_0)$, equipped with the product $[u] \ast [v] = \pi_0(u_0 v_0)$.

It is clear that the image of $v$ is precisely the zero mode of the corresponding field, modified to remove any (PBW-ordered) terms that annihilate all ground states.  The product $\ast$ is then just the product of the zero modes, with annihilating terms then removed.  The Ramond Zhu algebra $\rzhu{\VOA{V}_{\kk}}$ is obtained in exactly the same way, but restricted to the bosonic orbifold of $\VOA{V}_{\kk}$ (as fermionic fields will not have zero modes in the Ramond sector).

Let $\kk = -\frac{3}{2} + \frac{p}{2v}$, with the restrictions on $p$ and $v$ given in \eqref{eq:admkosp}, and let $\chi_{p,v}$ denote the \sv{} of $\VOA{V}_{\kk}$ that generates the ideal by which one quotients to obtain $\ospmodel{p,v}$.  We have the following useful results.
\begin{proposition} \label{prop:zhuresults}
	\leavevmode
	\begin{enumerate}
		\item \label{it:zhuunivns} \cite[Lem.~2.1]{KacVer94} $\nszhu{\VOA{V}_{\kk}} \cong \oenvalg$.
		\item \label{it:zhuunivr} \cite[Prop.~6]{RidAdm17} $\rzhu{\VOA{V}_{\kk}} \cong \slenvalg$.
		\item \label{it:zhusvdim} \cite{KacStr79} $\chi_{p,v}$ has conformal dimension $\frac{1}{2} (p-1) v$ and $h_0$-charge $p-1$.
		\item \label{it:zhuminmod} \cite[Prop.~7]{RidAdm17} $\zhu{\bullet}{\ospmodel{p,v}} \cong \dfrac{\zhu{\bullet}{\VOA{V}_{\kk}}}{\ideal{[y_0^{p-1} \chi_{p,v}]}}$, for $\bullet = \NS,\ \R$.
		\item \label{it:zhuideal} \cite[Lem.~3.3]{WooAdm18} The (\ns{} and Ramond) Zhu ideals $\ideal{[y_0^{p-1} \chi_{p,v}]}$ are not zero.
	\end{enumerate}
\end{proposition}
\noindent In part \ref{it:zhuminmod}, we can always replace $y_0^{p-1} \chi_{p,v}$ by $\chi_{p,v}$ in the \ns{} case.  However, this is only valid in the Ramond case if $p$ is odd because, otherwise, $\chi_{p,v}$ is fermionic and so has no zero mode.  We remark that one can easily prove the \ns{} case of part \ref{it:zhuideal} in an identical fashion to the corresponding \ns{} proof for the $N=1$ superconformal minimal models, given in \cite[Lem.~4.6]{BloSVir16}.  Here, the Ramond proof follows from the \ns{} one by spectral flow (this proof is much more subtle in the $N=1$ case).

\subsection{Completeness when $v=1$} \label{sec:completev=1}

We suppose first that $v=1$, hence that $p = 2 \kk + 3$ is odd (and at least $3$), so that $\kk \in \ZZ_{\ge 0}$.  The \sv{} $\chi_{p,1}$ is therefore bosonic and thus its zero mode generates the ideal in \cref{prop:zhuresults}\ref{it:zhuminmod} (by the remark following it).  The corresponding field may be taken to have the form
\begin{equation}
	\chi_{p,1}(z) = \normord{e(z)^{\kk+1}},
\end{equation}
where $\normord{\cdots}$ denotes normal ordering.  Since the field involves no fermions, we may compute its zero mode in both the \ns{} and Ramond sectors by inductively using the standard formula for the modes of a normally ordered product of fields.  In both cases, the result is
\begin{equation}
	[\chi_{p,1}] = e^{\kk+1}.
\end{equation}
We mention that this calculation simplifies greatly because all the $e_m$ commute among themselves.

Consider now the \ns{} sector.  By \cref{thm:zhu} and \cref{prop:zhuresults}, parts \ref{it:zhuunivns} and \ref{it:zhuminmod}, we know that $\Mod{M}$ will be an irreducible \ns{} relaxed \hw{} $\ospmodel{p,1}$-module if and only if its space of ground states is an irreducible weight $\osp$-module annihilated by $e^{\kk+1} = x^{2(\kk+1)}$.  Because $x$ acts nilpotently, it follows that this ground state $\osp$-module is actually \hw{} and, by comparing with the irreducible \ns{} \hw{} $\aosp$-modules listed in \cref{sec:osp}, we conclude that $\Mod{M} = \aospfin{\lambda}$ or $\parrev \aospfin{\lambda}$, for some $\lambda \le k$.  In the language of \cref{subsec:unitary}, we thereby obtain the irreducible \ns{} $\ospmodel{p,1}$-modules $\aospfin{r,0}$, with $r=1,3,\dots,2\kk+1=p-2$, and their parity reversals.  These are therefore the only irreducible \ns{} relaxed \hw{} $\ospmodel{p,1}$-modules, up to isomorphism.

Adapting this argument to the Ramond sector, we must replace \cref{prop:zhuresults}\ref{it:zhuunivns} by part \ref{it:zhuunivr}.  Thus, $\Mod{M}$ is an irreducible relaxed \hw{} $\ospmodel{p,1}$-module if and only if its space of ground states is an irreducible weight $\sltwo$-module annihilated by $e^{\kk+1}$.  Again, this means that the ground state module is \hw{} and an otherwise identical analysis concludes that the only irreducible Ramond relaxed \hw{} $\ospmodel{p,1}$-modules are the $\aospfin{r,0}$, with $r=2,4,\dots,2\kk+2=p-1$, and their parity reversals, again up to isomorphism.

We remark that the preceding analysis did not actually make use of the explicit constructions of irreducible $\ospmodel{p,v}$-modules reported in \cref{subsec:unitary}.  Nevertheless, it is worth pointing out that we now know, after performing the Zhu analysis, that our constructions resulted in a complete set of irreducibles, for each $p \in 2\ZZ+3$, up to parity.

\subsection{Completeness for general $v$} \label{sec:completev}

For general $v$, the direct classification argument used in the previous \lcnamecref{sec:completev=1} becomes much more difficult because an explicit formula for the \sv{} $\chi_{p,v}$ is not so easily determined.  (An implicit formula in terms of symmetric functions is used in \cite{WooAdm18}; we expect that it may also be possible to use the implicit Malikov-Feigin-Fuchs formula \cite{MalSin86} as well.)  We therefore describe a different approach that relies on the fact that we have already constructed many irreducible $\ospmodel{p,v}$-modules.  Our strategy is to show that the existence of any additional $\ospmodel{p,v}$-modules would violate a bound that we derive from Zhu considerations, thereby proving completeness.

We first analyse the \ns{} sector.  By \cref{prop:zhuresults}\ref{it:zhuunivns} and \ref{it:zhuminmod}, the Zhu ideal is generated by the image $[\psi]$ of the state $\psi = y_0^{p-1} \chi_{p,v}$ in $\nszhu{\VOA{V}_{\kk}} \cong \oenvalg$.  This image is not zero, by \cref{prop:zhuresults}{\ref{it:zhuideal}.  By considering the conformal dimension of $\psi$, we deduce that the number of modes in each of its PBW-monomials cannot exceed $\frac{1}{2} (p-1) v$ (\cref{prop:zhuresults}\ref{it:zhusvdim}), hence the same must be true for $\psi_0$ (using the standard normal ordering formulae) and $[\psi]$.  On the other hand, $\psi$ has $h_0$-charge $0$, so $[\psi] \in \oenvalg$ may be expressed \cite[Sec.~2.2]{RidAdm17} as a non-zero polynomial $P(h,\scas)$ in $h$ and the super-Casimir $\scas$.  We assign degrees $1$ to both $h$ and $\scas$ so as to get a bound on the \emph{total degree} of $P$:
\begin{equation} \label{eq:tdegP}
	\tdeg P \le \frac{(p-1)v}{2}.
\end{equation}

The reason for $\deg \scas = 1$ is a little subtle.  Na\"{\i}vely, one would think this degree should be $2$ as $\scas$ is quadratic in $x$ and $y$, see \eqref{eq:defscas}.  The point here is that we are not grading $\oenvalg$, but just the part of $h_0$-charge $0$.  So, while $xy = \frac{1}{2} (h+\scas - \frac{1}{2})$ suggests that $\scas$ should be degree $2$, because two modes results in one $\scas$, we also have
\begin{equation}
	ef = -x^2 y^2 = -\frac{1}{2} x \brac*{h+\scas - \frac{1}{2}} y = -\frac{1}{4} \brac*{h+\scas - \frac{1}{2}} \brac*{h-\scas - \frac{3}{2}},
\end{equation}
which makes it clear that two modes can result in a $\scas^2$.  Thus, $\deg \scas = 1$ is the correct choice.

As we have constructed the \ns{} $\ospmodel{p,v}$-modules $\aosprel{\lambda}{r,s}$ and $\parrev\aosprel{\lambda}{r,s}$, for an infinitude of $\lambda$ and all $r=1,\dots,p-1$ and $s=1,\dots,v-1$ with $r+s$ odd, their ground states must be annihilated by $\psi_0$, hence we have $P(\lambda,\pm\Sigma_{r,s}) = 0$ (see \eqref{eq:defSigma} for the definition of $\Sigma_{r,s}$).  Considering $P$ as a function of $\lambda$ alone (so holding $r$ and $s$ constant), this becomes $P(h,\pm\Sigma_{r,s}) = 0$ which implies that $\scas \pm \Sigma_{r,s}$ is a factor of $P(h,\scas)$ for all $r$ and $s$ in the above range.  Now, $\Sigma_{r,s} \neq 0$ in this range, unless $p$ and $v$ are even and $(r,s)=(\frac{p}{2},\frac{v}{2})$ (so $r+s=\frac{p+v}{2}$ is odd).  We may therefore write
\begin{equation} \label{eq:factorP}
	P(h,\scas) =
	\begin{cases*}
		Q(h,\scas) \prod \brac*{\scas^2 - \Sigma_{r,s}^2}, & if $p$ is odd, \\
		Q(h,\scas) \scas \prod \brac*{\scas^2 - \Sigma_{r,s}^2}, & if $p$ is even,
	\end{cases*}
\end{equation}
for some polynomial $Q$, where the products are over the pairs $(r,s)$ that give \emph{distinct non-zero} values of $\Sigma_{r,s}^2$.  It is easy to check that $\Sigma_{r,s}^2 = \Sigma_{r',s'}^2$ if and only if $(r',s')=(r,s)$ or $(p-r,v-s)$.  Moreover, $(r,s) = (p-r,v-s)$ is excluded as it gives $\Sigma_{r,s}=0$.  A careful count therefore shows that the number of pairs contributing to the products in \eqref{eq:factorP} is $\frac{1}{4} (p-1)(v-1)$, if $p$ is odd, and $\frac{1}{4} \sqbrac[\big]{(p-1)(v-1)-3}$, if $p$ is even.  It now follows from \eqref{eq:tdegP} and \eqref{eq:factorP} that
\begin{equation} \label{eq:Qbound}
	\tdeg Q \le
	\begin{cases*}
		\frac{p-1}{2}, & if $p$ is odd, \\
		\frac{p}{2}, & if $p$ is even.
	\end{cases*}
\end{equation}

We next use the fact that \ns{} $\ospmodel{p,v}$-modules $\aospfin{r,0}$ and $\parrev \aospfin{r,0}$, with $r=1,\dots,p-1$ odd, have also been constructed.  (The construction of the $\aospdispm{r,s}$ and $\parrev \aospdispm{r,s}$ does not help because their ground states have $\scas$-eigenvalues $\pm \Sigma_{r,s}$.)  Using \eqref{eq:defscas} or \eqref{eq:defSigma}, we find that their ground state $\scas$-eigenvalues are distinct, being of the form $\pm \frac{r}{2}$, and that they never coincide with any of the $\Sigma_{r',s'}$ with $r'=1,\dots,p-1$, $s'=1,\dots,v-1$ and $r'+s'$ odd --- the only possibility occurs when $p$ is even and $(r',s')=(r+\frac{p}{2},\frac{v}{2})$, but then $r'+s'$ is even.  These ground states therefore do not give zeroes of the products in \eqref{eq:factorP}, so they must give zeroes of $Q$.  In particular, the annihilation by $\psi_0$ of the \hwvs{} of the $\aospfin{r,0}$ and $\parrev \aospfin{r,0}$, which have $h_0$-charge $\lambda_{r,0} = \frac{1}{2} (r-1)$, leads to
\begin{equation}
	R_{\pm}(r) = Q \brac*{\frac{r-1}{2},\pm \frac{r}{2}} = 0, \qquad r=1,\dots,p-1\ \text{odd}.
\end{equation}
We conclude that
\begin{equation} \label{eq:Qbound'}
	\tdeg Q \ge \deg R_{\pm} \ge
	\begin{cases*}
		\frac{p-1}{2}, & if $p$ is odd, \\
		\frac{p}{2}, & if $p$ is even,
	\end{cases*}
\end{equation}
hence that the inequalities in \eqref{eq:Qbound} and \eqref{eq:Qbound'} are actually equalities.

This allows us to finally prove the completeness of the set of (isomorphism classes of) irreducible \ns{} relaxed \hw{} $\ospmodel{p,v}$-modules constructed in \cref{subsec:nonunimod}.  Any irreducible \ns{} relaxed \hw{} $\ospmodel{p,v}$-module $\Mod{M}$ is, \emph{a priori}, an $\aosp$-module, so must be one of those introduced in \cref{sec:osp}.  If $\Mod{M}$ is one of the ${}^{\NS} \aospdispm{\lambda}$, ${}^{\NS} \aosprel{\lambda}{\Sigma}$ or their parity reversals, then its bosonic ground states describe zeroes of $P(h,\scas)$ for infinitely many distinct $h_0$-charges.  As the $\scas$-eigenvalue of these states must all be the same, we must have $\Sigma = \pm \Sigma_{r,s}$, for some $r=1,\dots,p-1$, $s=1,\dots,v-1$ with $r+s$ odd, because $Q(h,\scas)$ cannot have infinitely many $h$-roots.  $\Mod{M}$ is thus one of the modules that we have constructed.

Alternatively, if $\Mod{M}$ is one of the ${}^{\NS} \aospfin{\lambda}$, with $\lambda \in \ZZ_{\ge 0}$, or their parity reversals, then its \hwv{} likewise describes a zero of $P(h,\scas)$.  Its $h_0$-charge is $\lambda$ and its $\scas$-eigenvalue is $\lambda + \frac{1}{2} \in \ZZ+\frac{1}{2}$ which never coincides with any of the $\pm\Sigma_{r,s}$ --- again, the only possible solution is $(r,s) = (\frac{p}{2}+2\lambda+1,\frac{v}{2})$, but then $r+s$ is even.  To get a zero of $P$, we must therefore have $R_{\pm}(2\lambda+1) = Q(\lambda,\pm(\lambda+\frac{1}{2})) = 0$.  However, we know the degree of $R_{\pm}$ and that all its roots correspond to $\lambda = \lambda_{r,0}$, for $r=1,\dots,p-1$ odd.  Thus, $\Mod{M}$ must likewise be one of the modules we have constructed.  The proof for the \ns{} sector is complete.

The proof in the Ramond sector is almost identical, so we only comment on the numerology and leave the details to the reader.  First, $[\psi]$ is now an $h_0$-charge $0$ element of $\rzhu{\VOA{V}_{\kk}} \cong \slenvalg$, hence $[\psi] = P(h,\cas)$ for some polynomial $P$, where $\cas$ is the $\SLA{sl}{2}$ Casimir \eqref{eq:defslcas}.  We again have \eqref{eq:tdegP}, though this requires consideration of generalised commutation relations in place of the usual prescription for normal ordering, with $\deg h = 1$ and $\deg \cas = 2$.  The existence of infinitely many Ramond $\ospmodel{p,v}$-modules $\aosprel{\lambda}{r,s}$, with $r+s$ even, now implies that $P$ decomposes as some polynomial $Q$ times a product of factors $(\cas - q_{r,s})$, where $q_{r,s}$ was defined in \eqref{eq:defq}.  The number of factors is the number of distinct values that $q_{r,s}$ takes: $\frac{1}{4} (p-1)(v-1)$, if $p$ is odd, and $\frac{1}{4} \sqbrac[\big]{(p-1)(v-1)+1}$, if $p$ is even.  This then gives upper bounds on $\tdeg Q$, being $\frac{p-1}{2}$, if $p$ is odd, and $\frac{p-2}{2}$, if $p$ is even.  Again, consideration of the $\aospfin{r,0}$, with $r$ even, saturates these bounds and the rest of the proof follows as before.

We conclude with two comments.  First, the completeness proof given here also works for $v=1$ where we only have the $\aospfin{r,0}$.  We presented the more direct $v=1$ proof in \cref{sec:completev=1} primarily to illustrate how easy it is.  Second, we mention that for general $v$, it is actually fairly straightforward now to completely identify generators of the \ns{} and Ramond Zhu ideals using $\chi_{p,v}$ and \cite[Lem.~3.4]{WooAdm18}.  In this way, we can derive explicit presentations for both Zhu's algebras, thereby arriving at a relatively painless proof of \cite[Thm.~3.6]{WooAdm18}.

\section{Fusion} \label{sec:Fusion}

\subsection{Grothendieck fusion rules for $\ospmodel{p,v}$} \label{sec:grfrosp}

One of the most convenient ways to compute the fusion rules of a rational bosonic \cft{} involves substituting its S-matrix entries into the Verlinde formula for fusion coefficients.  For fermionic theories, one can derive variations of the Verlinde formula as in \cite{EhoFus94,CanFusII15}.  For certain non-rational theories, there is a generalisation called the standard Verlinde formula \cite{CreLog13,RidVer14} that is conjectured to give the Grothendieck fusion coefficients of the theory, these being the structure constants of the Grothendieck group of the fusion ring.  We recall that the Grothendieck group is defined to be the $\ZZ$-span of the isomorphism classes of the irreducibles and that the image of a module in the Grothendieck group is the sum of the isomorphism classes of its composition factors.  A fermionic version of the standard Verlinde formula was recently tested successfully in \cite{RidAdm17} for the $\osp$ minimal model $\ospmodel{2,4}$.  We are thus confident that their result may be generalised straightforwardly to $\ospmodel{p,v}$ using the (super)character formulae derived here and the known S-matrices of the Virasoro and $\sltwo$ minimal models \cite{CreMod13}.

We shall, however, present an alternative approach to computing the (Grothendieck) fusion rules using \cref{thm:induction}, the coset \eqref{coset} and the known (Grothendieck) fusion rules of the Virasoro and $\sltwo$ minimal models $\vmodel{p,u}$ and $\slmodel{u,v}$. We shall illustrate the idea by computing the fusion of $\aospfin{r,0}$ and $\aospdisp{r',s'}$.  Both of these modules are defined, see \eqref{typesind}, as inductions of $\vmodel{p,u}\otimes\slmodel{u,v}$-modules.  Thus,
\begin{equation}
	\aospfin{r,0} \fuse \aospdisp{r',s'}
	= \ind{(\virr{r,1} \otimes \aslfin{1,0})} \fuse \ind{(\virr{r',1} \otimes \asldisp{1,s'})}
	\cong \ind{\brac*{(\virr{r,1} \otimes \aslfin{1,0}) \fuse (\virr{r',1} \otimes \asldisp{1,s'})}},
\end{equation}
as induction is preserved by fusion (\cref{thm:induction}).  Using the Virasoro fusion rules \eqref{eq:frvir} and the $\slmodel{u,v}$ fusion rules \eqref{eq:frsl2Lx}, this becomes
\begin{align}
	\aospfin{r,0} \fuse \aospdisp{r',s'}
	&\cong \ind{\brac*{(\virr{r,1} \fuse \virr{r',1}) \otimes (\aslfin{1,0} \fuse \asldisp{1,s'})}}
	\cong \bigoplus_{r''=1}^{p-1} \vcoe{p}{r,r'}{r''} \ind{(\virr{r'',1} \otimes \asldisp{1,s'})} \notag \\
	&= \bigoplus_{r''=1}^{p-1} \vcoe{p}{r,r'}{r''} \aospdisp{r'',s'},
\end{align}
where we have identified the final induced module using \eqref{typesind}.  Note that with the dictionaries presented in \cref{subsec:unitary,subsec:nonunimod}, these fusion rules completely capture the module structure of the fusion products, including the sector and relative parities.

In an identical fashion, \cref{thm:induction} gives the following $\ospmodel{p,v}$ fusion rules: 
\begin{subequations} \label{eq:frospAx}
	\begin{align}
		\aospfin{r,0} \fuse \aospfin{r',0} &= \bigoplus_{r''=1}^{p-1} \vcoe{p}{r,r'}{r''} \aospfin{r'',0}, \label{AA} \\
		\aospfin{r,0} \fuse \aospdispm{r',s'} &= \bigoplus_{r''=1}^{p-1} \vcoe{p}{r,r'}{r''} \aospdispm{r'',s'}, \\
		\aospfin{r,0} \fuse \aosprel{\lambda'}{r',s'} &= \bigoplus_{r''=1}^{p-1} \vcoe{p}{r,r'}{r''} \aosprel{\lambda'}{r'',s'}.
	\end{align}
\end{subequations}
Because fusion respects parity reversal and should respect spectral flow \cite[Prop.~2.11 and Eq.~(3.6)]{LiPhy97},
\begin{equation} \label{eq:parrsffusion}
	\Mod{M} \fuse \parrev \Mod{N} \cong \parrev (\Mod{M} \fuse \Mod{N}) \cong \parrev \Mod{M} \fuse \Mod{N}, \qquad
	\Mod{M} \fuse \sfmodosp{\Mod{N}}{} \cong \sfmodosp{\Mod{M} \fuse \Mod{N}}{} \cong \sfmodosp{\Mod{M}}{} \fuse \Mod{N},
\end{equation}
these fusion rules imply many others.  We remark that the fusion rules of the rational $\osp$ minimal models $\ospmodel{p,1}$ are given by \eqref{AA} alone.

Unfortunately, a complete set of irreducible $\ospmodel{p,v}$ fusion rules cannot be obtained in this way because the required $\slmodel{u,v}$ fusion rules are not known.  Instead, we have their Grothendieck versions \cite{CreMod13} which are reproduced for convenience in \eqref{eq:grfrsl2}.  We shall denote the Grothendieck fusion operation by $\Grfuse$ and the image of a module $\Mod{M}$ in the Grothendieck fusion ring by $\tGr{\Mod{M}}$.

The fact that $\Grfuse$ is well defined is not at all obvious.  A sufficient condition for this is that fusing with any fixed module from our category is exact, meaning that it respects the exactness of sequences.  For rational theories, such as the $\ospmodel{p,1}$, this is a theorem in the formalism of Huang, Lepowsky and Zhang \cite{HuaLog10}.  However, for the $\ospmodel{p,v}$ with $v \neq 1$, we have to assume that fusion is exact on a suitable module category.  Granting this, it follows that the fusion and Grothendieck fusion products of two modules $\Mod{M}$ and $\Mod{N}$ are related by
\begin{equation} \label{eq:deffusion}
	\Gr{\Mod{M} \fuse \Mod{N}} = \Gr{\Mod{M}} \Grfuse \Gr{\Mod{N}}.
\end{equation}
(This is, in fact, how $\Grfuse$ is defined.)  The exactness assumption being made is strong, but is not expected to be problematic.  Unfortunately, tools to verify it seem to be out of reach at present.

In any case, taking Grothendieck images respects tensor products and induction, the latter because it is defined in terms of fusion, hence the methods that led to the fusion rules \eqref{eq:frospAx} apply equally well to Grothendieck fusion rules.  This procedure thus determines the Grothendieck fusion rules involving all the irreducible $\ospmodel{p,v}$-modules of \eqref{typesind}.  Those that are not just the Grothendieck images of \eqref{eq:frospAx} (or its parity-reversed and spectral-flowed versions) are:
\begin{subequations}
	\begin{align}
		\Gr{\aospdisp{r,s}}\Grfuse\Gr{\aospdisp{r',s'}}&=
		\begin{dcases*}
			\sum_{r'',s''}{\vcoe{p,v}{(r,s),(r',s')}{(r'',s'')}}\left(\Gr{\sfmod{\aosprel{\lambda_{1,s+s'+1}}{r'',s''}}{}}+\Gr{\aospdisp{r'',s+s'}}\right), & if $s+s'<v$, \\
			\sum_{r'',s''}{\vcoe{p,v}{(r,s+1),(r',s'+1)}{(r'',s'')}}\left(\sfsymb\Gr{\aosprel{\lambda_{1,s+s'+1}}{r'',s''}}+\sfsymb^2\Gr{\aospdisp{r'',2v-2-s-s'}}\right), & if $s+s'\ge v$,
		\end{dcases*} \\
		\Gr{\aospdisp{r,s}}\Grfuse\Gr{\aosprel{\lambda'}{r',s'}}&=\sum_{r'',s''}{\vcoe{p,v}{(r,s+1),(r',s')}{(r'',s'')}}\Gr{\aosprel{\lambda'+\lambda_{1,s}}{r'',s''}}+\sum_{r'',s''}{\vcoe{p,v}{(r,s),(r',s')}{(r'',s'')}}\sfsymb\Gr{\aosprel{\lambda'+\lambda_{1,s+1}}{r'',s''}}, \\
		\Gr{\aosprel{\lambda}{r,s}}\Grfuse\Gr{\aosprel{\lambda'}{r',s'}}&=\sum_{r'',s''}{\vcoe{p,v}{(r,s),(r',s')}{(r'',s'')}}\left(\sfsymb\Gr{\aosprel{\lambda+\lambda'-k}{r'',s''}}+\Gr{\aosprel{\lambda+\lambda'+k}{r'',s''}}\right)\notag\\
		&\hspace{15mm}+\sum_{r'',s''}\left({\vcoe{p,v}{(r,s),(r',s'-1)}{(r'',s'')}}+{\vcoe{p,v}{(r,s),(r',s'+1)}{(r'',s'')}}\right)\Gr{\aosprel{\lambda+\lambda'}{(r'',s'')}}.
	\end{align}
\end{subequations}
Here, the sums over $r''$ always run from $1$ to $p-1$ while the sums over $s''$ always run from $1$ to $v-1$.  These fusion rules can be extended to include parity reversals and spectral flows using the Grothendieck versions of \eqref{eq:parrsffusion}.

\subsection{Projective modules}\label{sec:projective}

In \cref{app:frsl2}, we conjecture structures, in the form of Loewy diagrams, for the staggered modules $\aslproj{r,s}^{\pm}$ of $\slmodel{u,v}$ (recalling that these are indecomposable modules on which $L^{\supsl}_0$ acts non-semisimply \cite{RidSta09,CreLog13}).  We also conjecture that they are projective.  Note that projective modules induce to projective modules.\footnote{The reason for this is that projectivity is preserved by any functor (here, induction: $\ind{}$) that is left adjoint to an exact functor (here, restriction: $\res{}$). This adjointness was proved in \cite[Lem.~2.61]{CreTen17} and the exactness of the restriction functor is easy to verify.}  We may therefore immediately lift these conjectures to $\ospmodel{p,v}$. The lifts of these proposed projective modules will be denoted by $\aospproj{r, s}^\pm$ and are defined by
\begin{gather}
\aospproj{r,s}^\pm=\ind{(\virr{r,1}\otimes\aslproj{1,s}^\pm)},\qquad 1\leq r \leq p-1\ \text{and}\ 0\leq s \leq v-1.
\end{gather}
Their restrictions are then
\begin{equation} \label{eq:genprojbr}
\res{\aospproj{r,s}^\pm} \cong \bigoplus_{i=1}^{u-1} \virr{r,i}\otimes\aslproj{i,s}^\pm
\end{equation}
and the corresponding Loewy diagrams take the form
\begin{equation} \label{eq:loewyosp}
			\begin{tikzpicture}[->,thick,>=latex,baseline=(c.base)]
				\node (top) at (0,2) {$\aospdispm{r,s}$};
				\node (left) at (-2,0) {$\sfmodosp{\aospdispm{r,s-1}}{-1}$};
				\node (right) at (2,0) {$\sfmodosp{\aospdispm{r,s+1}}{}$};
				\node (bottom) at (0,-2) {$\aospdispm{r,s}$};
				\draw (top) -- (left);
				\draw (top) -- (right);
				\draw (left) -- (bottom);
				\draw (right) -- (bottom);
				\node[nom] (c) at (0,0) {$\aospproj{r,s}^{\pm}$};
			\end{tikzpicture}
			\qquad \text{($s=0,1,\dots,v-1$).}
		\end{equation}
where we have introduced the following convenient notation:
\begin{equation} \label{eq:ospconvenientnotation}
	\aospdispm{r,-1} = \aospdismp{r,1}, \qquad
	\aospdisp{r,0} \equiv \aospfin{r,0} \equiv \aospdism{r,0} \qquad \text{and} \qquad
	\aospdispm{r,v} = \sfmodosp{\aospdispm{u-r,1}}{\pm 1}.
\end{equation}
Of course, the $\aospproj{r,s}^{\pm}$ are staggered and are expected to be projective.  There are also analogous statements obtained by applying parity reversal.

For completeness, we also lift the conjectured $\slmodel{u,v}$ fusion rules \eqref{fr:ExEsl2} to $\ospmodel{p,v}$ fusion rules in order to show how the $\aospproj{r,s}^{\pm}$ arise.  Let $\lambda \neq \xi^{\pm}_{1,1} \pmod{2}$ and $\mu \neq \xi^{\pm}_{r,s} \pmod{2}$, where we recall the definition in \eqref{eq:translation}.  Then, for all $1 \le r \le p-1$ and $2 \le s \le v-2$ (which requires that $v \ge 4$), we have the fusion rules
	\begin{equation} \label{fr:ExEosp}
		\aosprel{\lambda}{1,1} \fuse \aosprel{\mu}{r,s} =
		\begin{cases*}
			\aospproj{r,s-1}^+ \oplus \sfmodosp{\aosprel{\lambda+\mu+\kk}{r,s}}{-1} \oplus \aosprel{\lambda+\mu}{r,s+1}, & if $\lambda+\mu=-\frac{p+v}{2v} (s-1)$, \\
			\aospproj{u-r,v-s-1}^+ \oplus \sfmodosp{\aosprel{\lambda+\mu+\kk}{r,s}}{-1} \oplus \aosprel{\lambda+\mu}{r,s-1}, & if $\lambda+\mu=\frac{p+v}{2v} (s+1)$, \\
			\aospproj{u-r,v-s-1}^- \oplus \sfmodosp{\aosprel{\lambda+\mu-\kk}{r,s}}{} \oplus \aosprel{\lambda+\mu}{r,s-1}, & if $\lambda+\mu=-\frac{p+v}{2v} (s+1)$, \\
			\aospproj{r,s-1}^- \oplus \sfmodosp{\aosprel{\lambda+\mu-\kk}{r,s}}{} \oplus \aosprel{\lambda+\mu}{r,s+1}, & if $\lambda+\mu=\frac{p+v}{2v} (s-1)$, \\
			\sfmodosp{\aosprel{\lambda+\mu-\kk}{r,s}}{} \oplus \sfmodosp{\aosprel{\lambda+\mu+\kk}{r,s}}{-1} \oplus \aosprel{\lambda+\mu}{r,s-1} \oplus \aosprel{\lambda+\mu}{1,s+1}, & otherwise,
		\end{cases*}
	\end{equation}
	where $\lambda+\mu$ is always understood$\pmod{2}$.

\appendix

\section{Grothendieck fusion rules for the $\sltwo$ minimal models} \label{app:frsl2}

The Grothendieck fusion rules for the non-unitary minimal model $\slmodel{u,v}$ were computed in \cite{CreMod13} using the conjectural standard Verlinde formula of \cite{CreMod13,RidVer14}. The fusion rules of type $\aslfin{r,0}\fuse \aslfin{r',0}$ were recently proven in \cite{CreBra17} and confirm the Verlinde conjectures.
 The results, which were shown to be consistent with the irreducible fusion rules of \cite{GabFus01}, for $(u,v)=(2,3)$ (see \cite{CreMod12} for some corrections), and \cite{RidFus10}, for $(u,v)=(3,2)$, are recorded in the following \lcnamecref{grfusesl2}.
\begin{conjecture}[\cite{CreMod13}] \label{grfusesl2}
	The Grothendieck fusion rules of the irreducible relaxed \hw{} $\slmodel{u,v}$-modules satisfy
	\begin{equation} \label{eq:sfgrfusesl2}
		\Gr{\sfmodsl{\Mod{M}}{m}} \Grfuse \Gr{\sfmodsl{\Mod{N}}{n}} = \sfmodsl{\Gr{\Mod{M}} \Grfuse \Gr{\Mod{N}}}{m+n}.
	\end{equation}
	The ``non-spectrally flowed'' rules are as follows:
	\begin{subequations} \label{eq:grfrsl2}
		\begin{align}
			\Gr{\aslfin{r,0}}\Grfuse \Gr{\aslfin{r',0}}&=\sum_{r''}{\vcoe{u}{r,r'}{r''}}\Gr{\aslfin{r'',0}},\\
			\Gr{\aslfin{r,0}}\Grfuse\Gr{\asldisp{r',s'}}&=\sum_{r''}{\vcoe{u}{r,r'}{r''}}\Gr{\asldisp{r'',s'}},\\
			\Gr{\aslfin{r,0}}\Grfuse\Gr{\aslrel{\lambda'}{r',s'}}&=\sum_{r''}{\vcoe{u}{r,r'}{r''}}\Gr{\aslrel{\lambda'+r-1}{r'',s'}},\\
			\Gr{\asldisp{r,s}}\Grfuse\Gr{\asldisp{r',s'}}&=
			\begin{dcases*}
				\sum_{r'',s''}{\vcoe{u,v}{(r,s),(r',s')}{(r'',s'')}}\Gr{\sfsymbsl\left(\aslrel{\lambda_{r'',s+s'+1}}{r'',s''}\right)}\\
				\mspace{50mu}+\sum_{r''}{\vcoe{u}{r,r'}{r''}}\Gr{\asldisp{r'',s+s'}}, & \text{if} $s+s'<v$,\\
				\sum_{r'',s''}{\vcoe{u,v}{(r,s+1),(r',s'+1)}{(r'',s'')}}\Gr{\sfsymbsl\left(\aslrel{\lambda_{r'',s+s'+1}}{r'',s''}\right)}\\
				\mspace{50mu}+\sum_{r''}{\vcoe{u}{r,r'}{r''}}\Gr{\sfsymbsl\left(\asldisp{u-r'',s+s'-v+1}\right)}, & \text{if} $s+s'\ge v$,
			\end{dcases*} \\
			\Gr{\asldisp{r,s}}\Grfuse\Gr{\aslrel{\lambda'}{r',s'}}&=\sum_{r'',s''}{\vcoe{u,v}{(r,s+1),(r',s')}{(r'',s'')}}\Gr{\aslrel{\lambda'+\lambda^{\supsl}_{r,s}}{r'',s''}}+\sum_{r'',s''}{\vcoe{u,v}{(r,s),(r',s')}{(r'',s'')}}\Gr{\sfsymbsl \brac[\big]{\aslrel{\lambda'+\lambda^{\supsl}_{r,s+1}}{r'',s''}}}, \\
			\Gr{\aslrel{\lambda}{r,s}}\Grfuse\Gr{\aslrel{\lambda'}{r',s'}}&=\sum_{r'',s''}{\vcoe{u,v}{(r,s),(r',s')}{(r'',s'')}}\left(\Gr{\sfsymbsl \brac*{\aslrel{\lambda+\lambda'-k}{r'',s''}}}+\Gr{\sfsymbsl^{-1} \brac*{\aslrel{\lambda+\lambda'+k}{r'',s''}}}\right)\notag\\
			&\hspace{15mm}+\sum_{r'',s''}\left({\vcoe{p,v}{(r,s),(r',s'-1)}{(r'',s'')}}+{\vcoe{p,v}{(r,s),(r',s'+1)}{(r'',s'')}}\right)\Gr{\aslrel{\lambda+\lambda'}{r'',s''}}. \label{grfr:ExEsl2}
		\end{align}
	\end{subequations}
	Here, the sums over $r''$ always run from $1$ to $u-1$ while the sums over $s''$ always run from $1$ to $v-1$.
\end{conjecture}
\noindent We refer to \eqref{vircoe} for the definition of the (Virasoro) fusion coefficients that appear.

The known fusion rules for $(u,v)=(2,3)$ and $(3,2)$ involve additional reducible, but indecomposable, $\slmodel{u,v}$-modules with four composition factors each.  They are examples of staggered modules, in the sense of \cite{RidSta09,CreLog13}, possessing a non-diagonalisable action of $L^{\supsl}_0$.  As such, they are responsible for the logarithmic nature of the corresponding \cfts{}.  We believe that these staggered modules are projective and are therefore the projective covers of their irreducible heads (in an appropriate category of $\slmodel{u,v}$-modules).  We record this belief as a formal \lcnamecref{loewysl2} below, extending it to all admissible levels.

For convenience, let us agree to the following notation:
\begin{equation} \label{eq:convenientnotation}
	\asldispm{r,-1} = \asldismp{r,1}, \qquad
	\asldisp{r,0} \equiv \aslfin{r,0} \equiv \asldism{r,0} \qquad \text{and} \qquad
	\asldispm{r,v} = \sfmodsl{\asldispm{u-r,1}}{\pm 1}.
\end{equation}
The projective covers of the $\asldispm{r,s}$, for $s=0,1,\dots,v-1$, shall be denoted by $\aslproj{r,s}^{\pm}$.  We shall sometimes drop the label $\pm$ when $s=0$ in accordance with the second identification of \eqref{eq:convenientnotation}.

The structures of the (conjectured) projective covers will be described in terms of their \emph{Loewy diagrams}.  This is a picture in which the composition factors of the module are arranged in horizontal layers.  The bottom layer contains the composition factors of the module's socle.  The next layer up contains the composition factors of the socle of the quotient of the module by its socle.  This continues up until we reach the top layer which contains the composition factors of the module's head.  Two composition factors in adjacent layers may be connected by an arrow if there is an indecomposable subquotient with only these as its composition factors.  In this case, the arrow points from the quotient to the submodule of the subquotient.  Roughly speaking, the arrows indicate the ``direction'' taken by the action of the algebra.  We refer to \cite[App.~A.4]{CreLog13} for an elementary introduction to Loewy diagrams that describes the idea in more detail.

With this background in place, we can now state our conjecture for the projective covers of the irreducible $\slmodel{u,v}$-modules.
\begin{conjecture} \label{loewysl2}
	\leavevmode
	\begin{itemize}
		\item The irreducible $\sfmodsl{\aslrel{\lambda}{r,s}}{\ell}$, with $\ell \in \ZZ$, $r=1,\dots,u-1$, $s=1,\dots,v-1$ and $\lambda \neq \lambda^{\supsl}_{r,s}, \lambda^{\supsl}_{u-r,v-s} \pmod{2}$, are projective and are hence their own projective covers.
		\item The Loewy diagram of the projective cover $\aslproj{r,s}^{\pm}$ of $\asldispm{r,s}$ is
		\begin{equation} \label{eq:loewysl2}
			\begin{tikzpicture}[->,thick,>=latex,baseline=(c.base)]
				\node (top) at (0,2) {$\asldispm{r,s}$};
				\node (left) at (-2,0) {$\sfmodsl{\asldispm{r,s-1}}{-1}$};
				\node (right) at (2,0) {$\sfmodsl{\asldispm{r,s+1}}{}$};
				\node (bottom) at (0,-2) {$\asldispm{r,s}$};
				\draw (top) -- (left);
				\draw (top) -- (right);
				\draw (left) -- (bottom);
				\draw (right) -- (bottom);
				\node[nom] (c) at (0,0) {$\aslproj{r,s}^{\pm}$};
			\end{tikzpicture}
			\qquad \text{($s=0,1,\dots,v-1$).}
		\end{equation}
	\end{itemize}
\end{conjecture}
\noindent The projective cover of $\sfmodsl{\asldispm{r,s}}{\ell}$ is then $\sfmodsl{\aslproj{r,s}^{\pm}}{\ell}$ and its Loewy diagram is obtained from that of $\asldispm{r,s}$ by applying $\sfsymbsl^{\ell}$ to each composition factor.  (Indeed, that of $\aslproj{r,v-1}^+$ is the image under $\sfsymbsl$ of that of $\aslproj{r,0}$.)  We remark that it is easy to prove that almost all of the $\sfmodsl{\aslrel{\lambda}{r,s}}{\ell}$ are projective.

Evidence for  the conjectured Loewy diagrams \eqref{eq:loewysl2} comes from trying to lift the Grothendieck fusion rules of \cref{grfusesl2} to actual fusion rules.  We expect that the physically consistent category of $\slmodel{u,v}$-modules should be, among other things, rigid and tensor.  The associative tensor product is, of course, fusion and rigidity ensures that fusing with any fixed module defines an exact functor on the category \cite[Prop.~4.2.1]{EtiTen15}.  This means that the Grothendieck group of the category inherits a well-defined product $\Grfuse$ from the fusion product $\fuse$, as in \eqref{eq:deffusion}.  Another consequence of rigidity is that the projectives of the category form a tensor ideal: the fusion product of a projective, in particular one of the irreducible $\aslrel{\lambda}{r,s}$, with any module is again projective \cite[Prop.~4.2.12]{EtiTen15}.

As the $\aslfin{r,0}$, $\asldispm{r,s}$ and $\aslrelredpm{r,s}$, along with their spectral flows, cannot be projective, there are not many ways to arrange the composition factors, obtained from \cref{grfusesl2}, of a fusion product involving an irreducible $\aslrel{\lambda}{r,s}$ so that the result could be projective.  Indeed, if we also insist on projectives being self-dual, a desirable property in view of the non-degeneracy of two-point correlation functions \cite{RidPer08}, then the arrangement is often essentially unique.  This is reflected in the following \lcnamecref{fusesl2} for a particular subset of the $\slmodel{u,v}$ fusion rules.
\begin{conjecture} \label{fusesl2}
	Let $\lambda \neq \lambda^{\supsl}_{1,1}, \lambda^{\supsl}_{u-1,v-1} \pmod{2}$ and $\mu \neq \lambda^{\supsl}_{r,s}, \lambda^{\supsl}_{u-r,v-s} \pmod{2}$.  Then, for all $1 \le r \le u-1$ and $2 \le s \le v-2$ (which requires that $v \ge 4$), we have the fusion rules
	\begin{equation} \label{fr:ExEsl2}
		\aslrel{\lambda}{1,1} \fuse \aslrel{\mu}{r,s} =
		\begin{cases*}
			\aslproj{r,s-1}^+ \oplus \sfmodsl{\aslrel{\lambda+\mu+\kk}{r,s}}{-1} \oplus \aslrel{\lambda+\mu}{r,s+1}, & if $\lambda+\mu=\lambda^{\supsl}_{r,s-1}$, \\
			\aslproj{u-r,v-s-1}^+ \oplus \sfmodsl{\aslrel{\lambda+\mu+\kk}{r,s}}{-1} \oplus \aslrel{\lambda+\mu}{r,s-1}, & if $\lambda+\mu=\lambda^{\supsl}_{u-r,v-s-1}$, \\
			\aslproj{u-r,v-s-1}^- \oplus \sfmodsl{\aslrel{\lambda+\mu-\kk}{r,s}}{} \oplus \aslrel{\lambda+\mu}{r,s-1}, & if $\lambda+\mu=\lambda^{\supsl}_{r,s+1}$, \\
			\aslproj{r,s-1}^- \oplus \sfmodsl{\aslrel{\lambda+\mu-\kk}{r,s}}{} \oplus \aslrel{\lambda+\mu}{r,s+1}, & if $\lambda+\mu=\lambda^{\supsl}_{u-r,v-s+1}$, \\
			\sfmodsl{\aslrel{\lambda+\mu-\kk}{r,s}}{} \oplus \sfmodsl{\aslrel{\lambda+\mu+\kk}{r,s}}{-1} \oplus \aslrel{\lambda+\mu}{r,s-1} \oplus \aslrel{\lambda+\mu}{r,s+1}, & otherwise,
		\end{cases*}
	\end{equation}
	where $\lambda+\mu$ is always understood$\pmod{2}$.

	When $s=1$ or $s=v-1$, these fusion rules are modified to remove any $\aslrel{\nu}{r,s'}$, with $s'=0$ or $v$, and remove any direct summands that do not appear in all expressions corresponding to the same value of $\lambda+\mu \pmod{2}$.  For example, the fusion rule for $s=1$, $v \ge 3$ and $\lambda+\mu = \lambda^\supsl_{r,0} \pmod{2}$ becomes
	\begin{equation}
		\aslrel{\lambda}{1,1} \fuse \aslrel{\mu}{r,1} = \aslproj{r,0} \oplus \aslrel{\lambda+\mu}{r,2},
	\end{equation}
	because $\lambda^\supsl_{r,0} = \lambda^\supsl_{u-r,v}$ and the spectrally flowed summands in the first and fourth cases of \eqref{fr:ExEsl2} are different.  When $v=2$, we would also have to remove the $\aslrel{\lambda+\mu}{r,2}$ from the \rhs.
\end{conjecture}
\noindent In fact, the Loewy diagrams \eqref{eq:loewysl2} were deduced by analysing the possible arrangements for the composition factors appearing in the Grothendieck counterpart \eqref{grfr:ExEsl2} (with $r,s=1$).  It is, of course, possible to similarly conjecture the remaining fusion rules involving the irreducible $\slmodel{u,v}$-modules.  These fusion rules will be reported in \cite{Tian}.

\flushleft

\providecommand{\opp}[2]{\textsf{arXiv:\mbox{#2}/#1}}
\providecommand{\pp}[2]{\textsf{arXiv:#1 [\mbox{#2}]}}

\end{document}